\documentclass[aip,rsi,amsmath,amssymb,reprint,%
]{revtex4-1}
\usepackage{graphicx}
\usepackage{ulem}
\usepackage{color}
\usepackage{amssymb}
\usepackage{subfigure}
\usepackage{siunitx} 

\makeatletter
\newcommand*{\addFileDependency}[1]{
\typeout{(#1)}
\@addtofilelist{#1}

\IfFileExists{#1}{}{\typeout{No file #1.}}
}\makeatother

\newcommand*{\myexternaldocument}[1]{%
\externaldocument{#1}%
\addFileDependency{#1.tex}%
\addFileDependency{#1.aux}%
}

\usepackage[utf8]{inputenc} 
\usepackage[T1]{fontenc}
\usepackage{bigstrut}
\usepackage{booktabs}
\usepackage{textgreek}
\usepackage{xr}

\usepackage[dvipsnames]{xcolor}

\myexternaldocument{SAM_SOM}

\begin{document}
\title{Local work-function manipulation by external optical stimulation}
\author{Jan B\"ohnke}
\author{Beatrice Andres}
\author{Larissa Boie}
\author{Angela Richter}
\author{Cornelius Gahl}
\author{Martin Weinelt}
\email[Corresponding author: ] {weinelt@physik.fu-berlin.de}
\affiliation{Freie Universit{\"a}t Berlin, Fachbereich Physik, Arnimallee 14, 14195 Berlin, Germany}
\author{Wibke Bronsch}
\email[Corresponding author: ] {wibke.bronsch@fu-berlin.de}
\altaffiliation[Current address: ]{Elettra - Sincrotrone Trieste S.C.p.A., Strada Statale 14 - km 163.5 in AREA Science Park, 34149 Basovizza, Trieste, Italy}
\affiliation{Freie Universit{\"a}t Berlin, Fachbereich Physik, Arnimallee 14, 14195 Berlin, Germany}
\date{\today}

\begin{abstract}
Strongly differing static dipole moments of the \textit{trans} and \textit{cis} isomers of photochromic azobenzene allow for optical switching of the work function of azobenzene-functionalized self-assembled monolayers (SAMs). We apply these properties in a fundamental experiment to manipulate the area size of the switched SAM. Azobenzene molecules were excited by ultraviolet laser illumination and the transient isomerization profile of the SAM was spatially resolved recording photoemission electron microscopy images. Thereby we demonstrate the spatial tuning of the SAM's work function and discuss the role of the laser spot profile in generating sharp edges or gradual changes of the work function. 
\end{abstract}

\maketitle

\begin{figure}[t]
        \includegraphics[width=9cm]{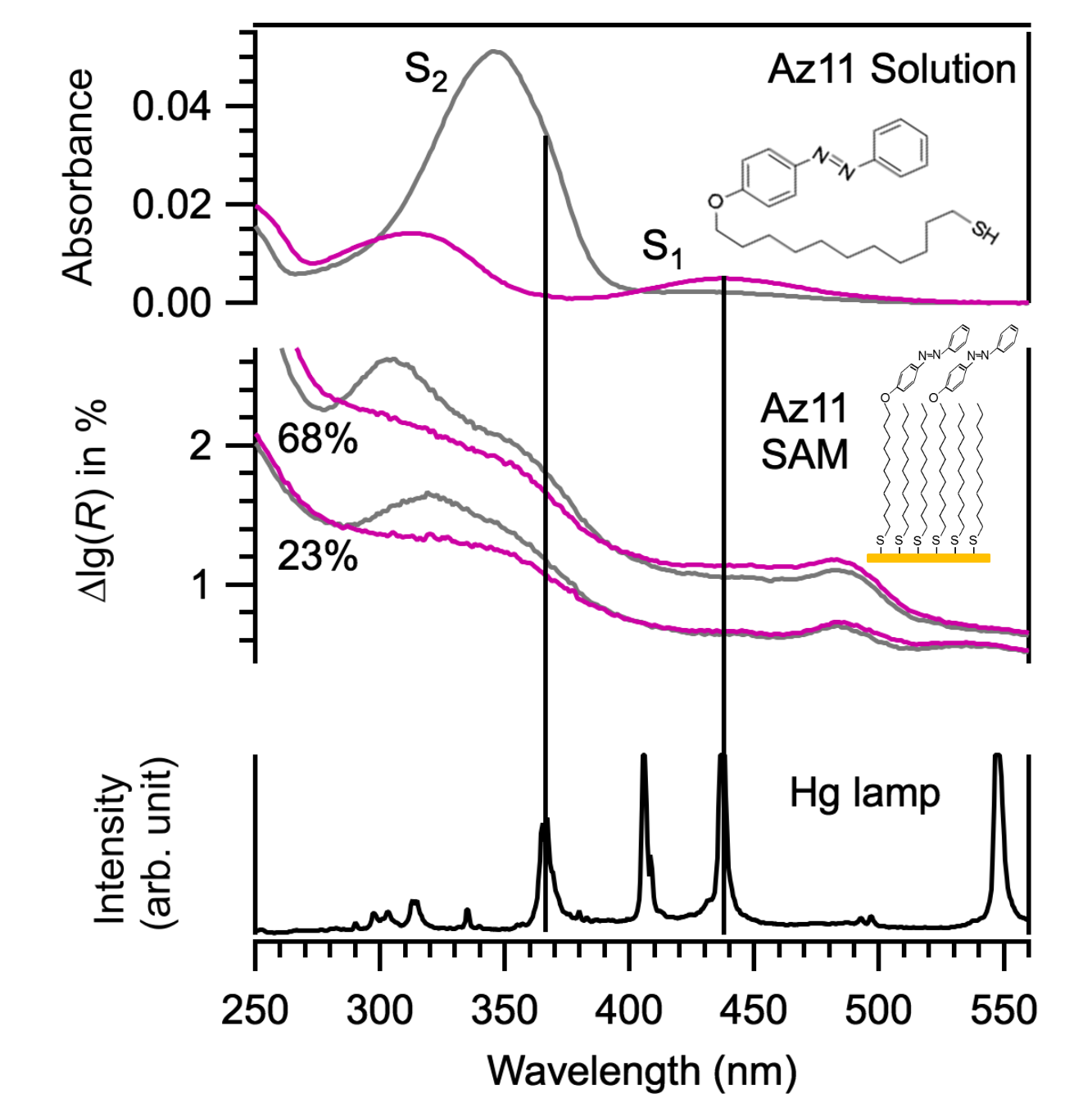}
        \caption{Comparison of the mercury lamp emission lines and the absorbance of Az11. Top: Absorbance of non-interacting \textit{trans} (grey curve) and {\it cis} (magenta curve) Az11 molecules in solution. Inset: Structural formula of Az11 in \textit{trans} configuration. Middle: Differential-reflectance signal of a 23 and 68\,\% Az11-SAM in an all-\textit{trans} configuration (grey curves) and under 365\,nm illumination (magenta curves) as published in Moldt {\it et al.}\cite{moldt2015} Inset: Cartoon of the Az11-SAM in \textit{trans} configuration diluted with C12 spacer-molecules. Bottom: Emission spectrum of the mercury lamp. Vertical lines indicate the most relevant emission lines of the Hg lamp in relation to the absorption bands of non-interacting and excitonically coupled Az11 molecules.
}
\label{fig:Hg_SAM}
\end{figure}

Azobenzene-based self-assembled monolayers (SAMs) are prototype systems for photoresponsive surfaces. \cite{steen2023,ichimura2000,crivillers2011a,ahqune2008,crivillers2011,crivillers2013,nagahiro2009}
Their photochromic switching ability can, \textit{e.g.}, be used to tune level alignments at interfaces.\cite{ahqune2008,crivillers2011,crivillers2013,nagahiro2009}
Here we discuss local changes of the SAM's work function induced by altering the static dipole moment of the chromophores with controlled photoisomerization. 
Azobenzene is a conformational switch, consisting of two phenyl rings coupled via a dinitrogen bond (see inset in Fig.~\ref{fig:Hg_SAM}). 
In its \textit{trans} form the molecule is planar and the two rings are on opposite sides of the dinitrogen bond axis. In the \textit{cis} conformation, both rings are on the same side of the reference plane. 
Therefore, the molecule has no static dipole moment in the \textit{trans} configuration, but is strongly polar in its \textit{cis} state.\cite{hartley1939} 
Dissolved in solution, light in the blue to ultraviolet (UV) range triggers photoswitching between the two isomers.\cite{bandara2012}
For fabricating photoresponsive surfaces we linked azobenzene to Au(111)/Mica substrates via an alkanethiol chain following an elaborated procedure.\cite{freyer2009,klajn2010,gahl2010,moldt2015,moldt2016}
Utilizing a linker with a chain length of 11 carbon atoms, the azobenzene head groups are sufficiently decoupled from the surface (see Fig.~\ref{fig:Hg_SAM} insets). 
We will denominate the azobenzene derivative 11-(4-(phenyldiazenyl)phenoxy)-undecane-1-thiol used in this work as Az11. 
It is essential to dilute the chromophore density to prevent steric hindrance.\cite{gahl2010,heinemann2012}
We observed efficient photoswitching in mixed SAMs of Az11 and a pure alkanethiolate of comparable length (C12) prepared in a wet-chemical process.\cite{moldt2015} 
In this work we used samples with \qty{50}{\percent} Az11 surface coverage.\cite{moldt2015}

Near edge X-ray absorption fine structure (NEXAFS) spectroscopy showed that in a pure Az11 SAM the {\it trans} chromophores are tilted by about \qty{30}{\degree} from the surface plane.\cite{gahl2010} 
In diluted SAMs the chromophores tilt further towards the SAM surface.\cite{moldt2016}
When switching into its {\it cis} form, the molecule flips the upper phenyl ring and the $z$-component of its static dipole moment points towards the surface. Thus the work function of the SAM/Au(111) sample increases with the number of molecules in \textit{cis} configuration.\cite{ahqune2008,nagahiro2009,bronsch2017} 
This allows us to locally tune the work function and image areas of switched molecules in the SAM via spatially resolved photoelectron emission. 
Although external work-function manipulations of azobenzene-functionalized surfaces have been studied by several groups \cite{ahqune2008,crivillers2011,crivillers2013,nagahiro2009,stiller1999,gustina1999,schuster2019}, so far no analysis of the switched area was performed. 
Controlling homogeneity and locality of the switched area is highly relevant for future applications.
Here we use the photoemission electron microscopy (PEEM) mode of a time-of-flight momentum microscope for getting access to spatial modulations in the \unit{\micro\meter} range. PEEM allows us to analyse the size and profile of switched areas. We demonstrate that the absolute value of the local work function and hence the isomerization profile as well as the time span to switch the work function can be tuned by laser illumination.\\ 

In Fig.~\ref{fig:Hg_SAM} we compare the absorbance of Az11 molecules in methanol solution and the differential reflectance of mixed SAMs from Ref.~\onlinecite{moldt2015} with the spectrum of a mercury arc lamp.
In the SAM, the S$_2$ band splits and its maximum shifts to higher energies with increasing Az11 percentage (Fig.\,\ref{fig:Hg_SAM}, middle panel). This is attributed to excitonic coupling between the {\it trans} isomers, whereby the $\pi\pi^{\ast}$ transition-dipole moments form an H-aggregate. 
Molecules switched to the \textit{cis} state show no excitonic coupling.\cite{bronsch2017a} 
Irradiation at the low-energy edge of the S$_2$ absorption band at about \qty{370}{nm} leads to {\it trans--cis} isomerization in the SAM. 
The {\it cis--trans} back-isomerization is triggered by illumination either at the high energy edge of the S$_2$ band or in the S$_1$ band, showing up at around \qty{440}{nm}.\cite{bronsch2017a}

We mainly used two complementary photon sources, a continuous wave (cw) laser (\qty{372}{nm}) and a Hg lamp for illumination in the S$_2$ and S$_1$ absorption bands, respectively. Simultaneous excitation facilitates optical tuning of the {\it cis--trans} ratio at the functionalized surface. 
The cw laser initiates the {\it trans--cis} isomerization only, whereas the Hg lamp contributes to both {\it trans--cis} and {\it cis--trans} configuration changes. 
As illustrated in Fig.~\ref{fig:Hg_SAM}, several emission lines of the mercury lamp are in the range of the S$_1$ and S$_2$ absorption bands of the Az11 SAM. 
The intense emission line at $\qty{436}{nm}$ illuminates in the S$_1$ band center, leading to efficient {\it cis--trans} isomerization. 
The emission line at \qty{365}{nm} instead triggers {\it trans--cis} switching.
Considering the relative photon flux of both emission lines we estimated the photostationary state (PSS) of the mercury lamp (Hg-PSS) as discussed in detail in the supplementary material (SM, Sec.\,\textrm{II}).
We find photon fluxes of $j_{\rm 436\,nm} = 1.9 \cdot 10^{14}$\,cm$^{-2}$ and $j_{\rm 365\,nm} = 3.3 \cdot 10^{14}$\,cm$^{-2}$ leading to a PSS with \qty{67}{\percent} {\it cis} molecules.
Starting from this PSS, a maximal work-function change of \qty{30}{meV} can be reached when tuning towards a pure {\it cis} SAM by simultaneous laser illumination at \qty{372}{nm}.\cite{bronsch2017a}

\begin{figure}[b]
        \includegraphics[width=8cm]{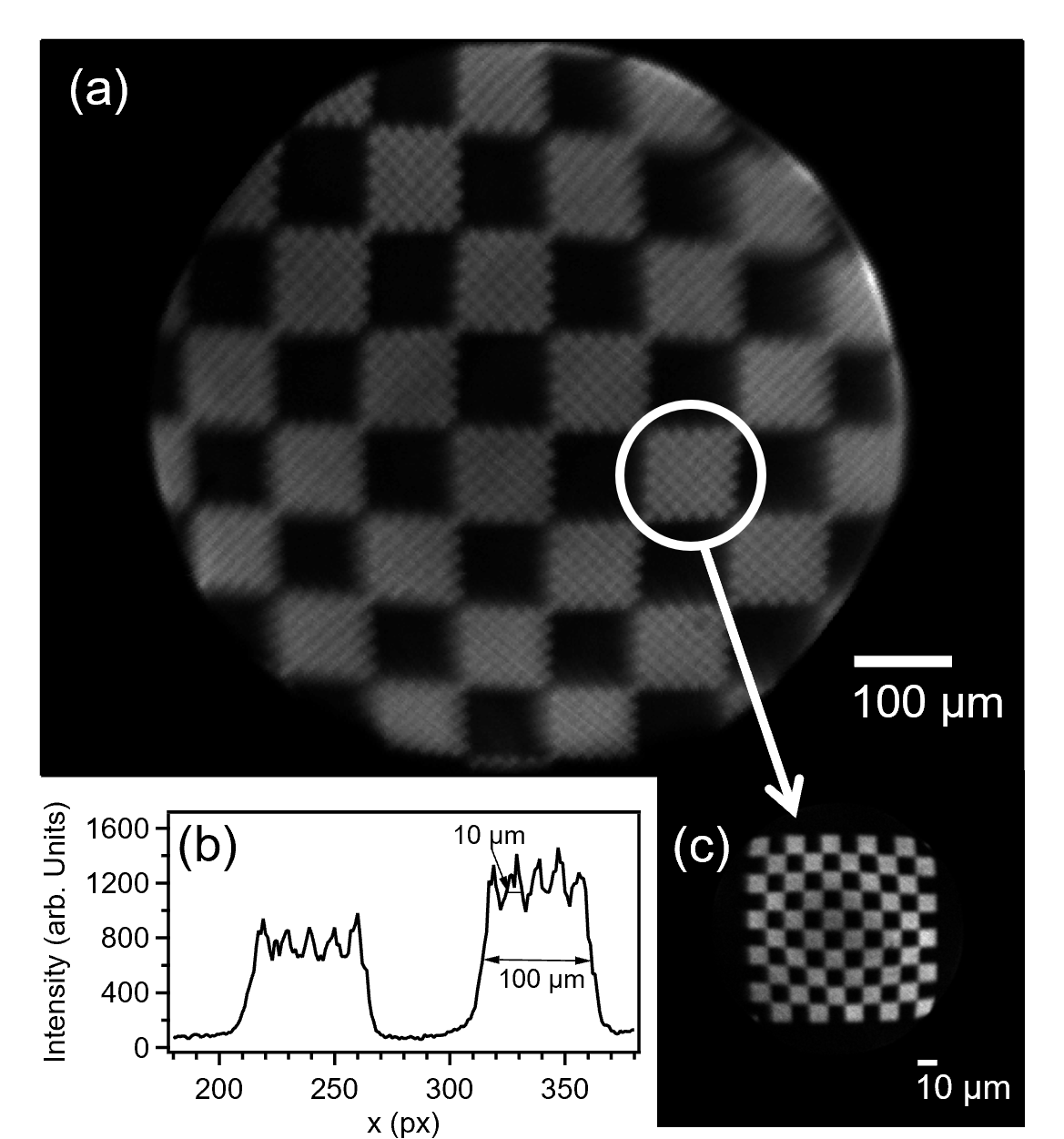}
        \caption{(a) Spatially-resolved photoelectron emission of a Chessy sample for calibration of the focusing and magnifying electron optics. We selected a PEEM FoV with a diameter of \qty{800}{\micro\meter} to spatially resolve the work-function changes. The line profile in $x$ direction in (b) indicates the ability to resolve \qty{10}{\micro\meter} squares with this setting. Inset (c) shows an image with a FoV of \qty{100}{\micro\meter} diameter and a spatial resolution below \qty{3}{\micro\meter} taken with a different lens magnification setting (see SM, Sec.\,\textrm{V}).}
\label{fig:Chessy}
\end{figure}

\begin{figure*}[t]
        \includegraphics{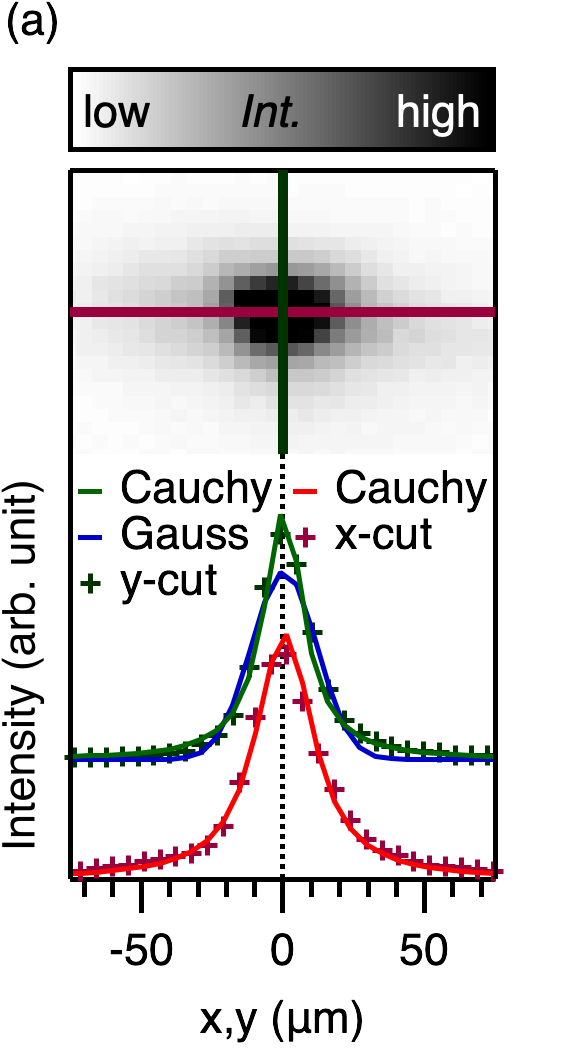}
        \includegraphics{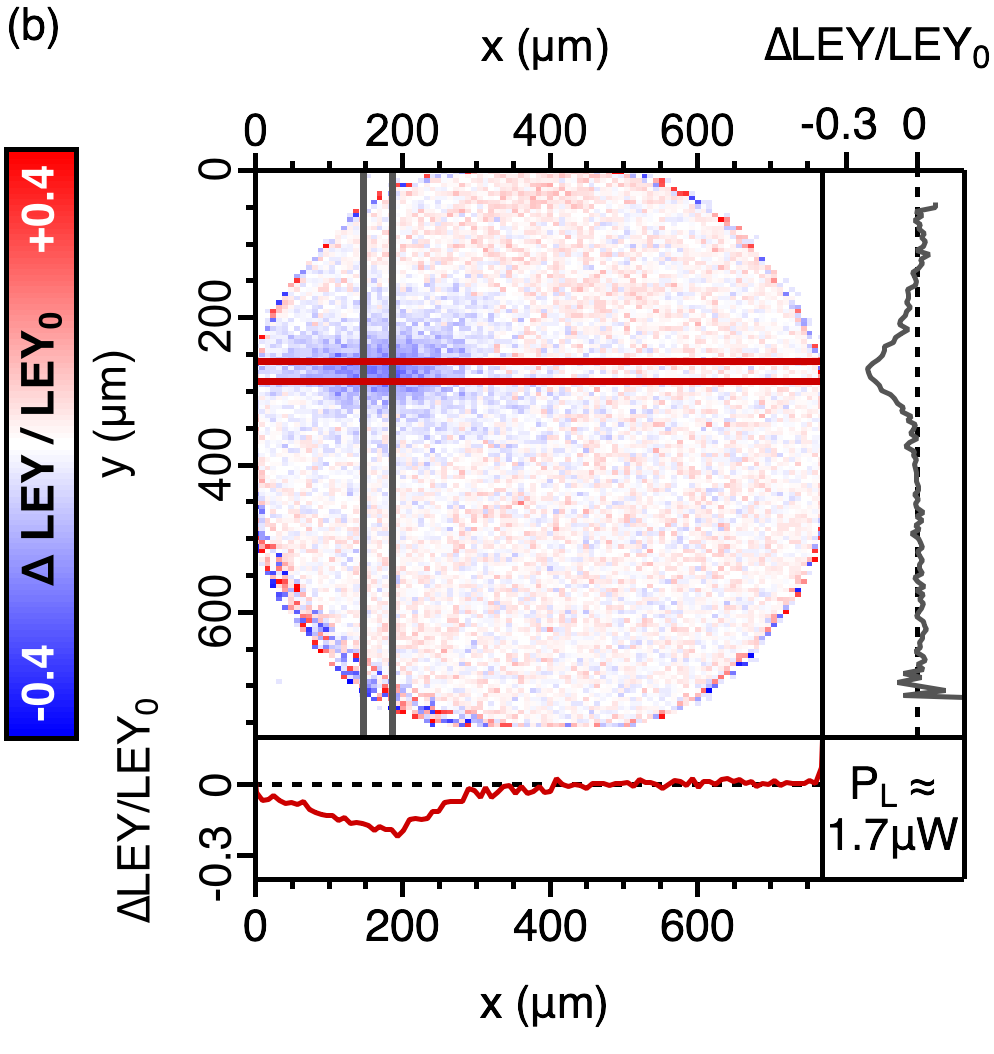}
        \includegraphics{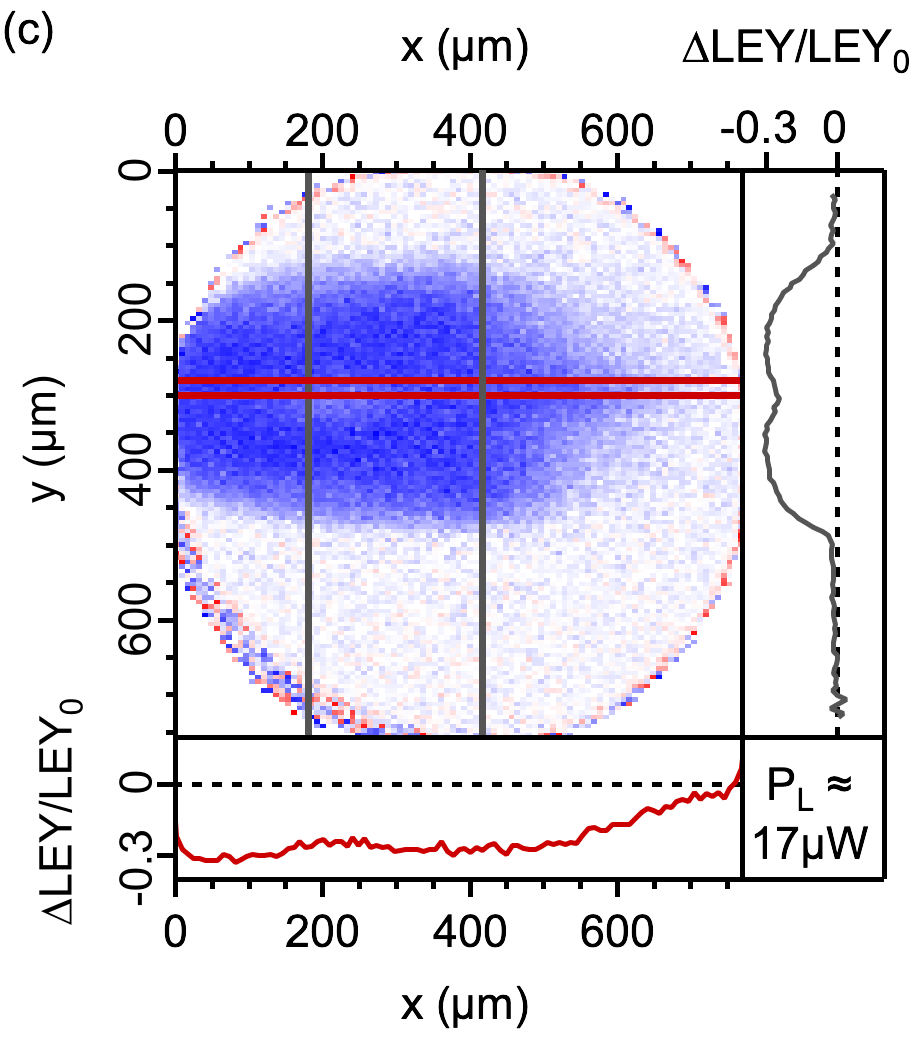}
           \caption{
        (a) CCD image of the cw laser spot. Vertical and horizontal cuts through the spot profile indicate contributions with a Lorentzian-like tail (Cauchy distribution). (b, c) Normalized PEEM images (LEY({\rm{Hg + 372}}) - LEY({\rm{Hg}}))/LEY({\rm{Hg}}) recorded for a laser power $P_{\rm{L}}$ of 1.7 and \qty{17}{\micro W}, respectively. The images show a decrease of the local electron yield (LEY) in the areas where the laser spot was positioned. Side panels show profiles in vertical and horizontal direction extracted in the range indicated by horizontal (grey) and vertical (red) lines.}
        \label{fig:PEEM}
\end{figure*}

The high-energetic lines of the mercury lamp lead to photoelectron emission from the functionalized surface.\cite{bronsch2017,bronsch2017a}
We perform spatially-resolved photoemission experiments in an ultra-high vacuum chamber using a momentum microscope.\cite{schonhense2015,schonhense2022} 
The instrument has been set up in collaboration with G.~Sch\"onhense. 
In this work we use the instrument's spatially resolving "Gaussian" imaging mode comparable to a photoelectron emission microscope.\cite{schonhense2015}
The field of view (FoV) on the sample surface is selected by real space apertures inserted between two zoom optics. 
Photoelectrons are detected after a time-of-flight tube by a delay line detector.\cite{oelsner2010}
A $(5 \times 5)$\,mm$^2$ chess-patterned gold-on-silicon sample (Chessy, Plano) was characterized for initial alignment and calibration of the focusing and magnifying electron optics, exploiting the instrumental resolution. 
Figure~\ref{fig:Chessy} shows spatially-resolved electron images of the $(10 \times 10)$\,\qty{}{\micro\meter}$^2$ chess patterns. With a FoV of \qty{100}{} and \qty{800}{\micro\meter} diameter, we obtained spatial resolutions below \qty{3}{} and \qty{16}{\micro\meter}, respectively (see SM, Sec.\,\textrm{V}). 
Since the measurements on the Az11 SAMs required a large FoV, we used the latter settings which still provide appropriate spatial resolution. 

{\it Trans--cis} isomerization of the SAM in a selected area was triggered focusing the \qty{372}{nm} laser beam onto the sample. An image of the laser spot recorded with a CCD camera is shown in Fig.~\ref{fig:PEEM}a together with horizontal ($x$) and vertical ($y$) beam profiles taken across the spot center. The focus has an elliptical profile with full width at half maximum (FWHM) of $19$ and $25 \pm 1 \, \unit{\micro\meter}$. 
Beam profiles were modeled using 2D Gaussian and Cauchy distributions. The latter accounts particularly better for the projection of the laser spot along the direction of incidence $x$. 

\begin{figure*}[t]
        \includegraphics{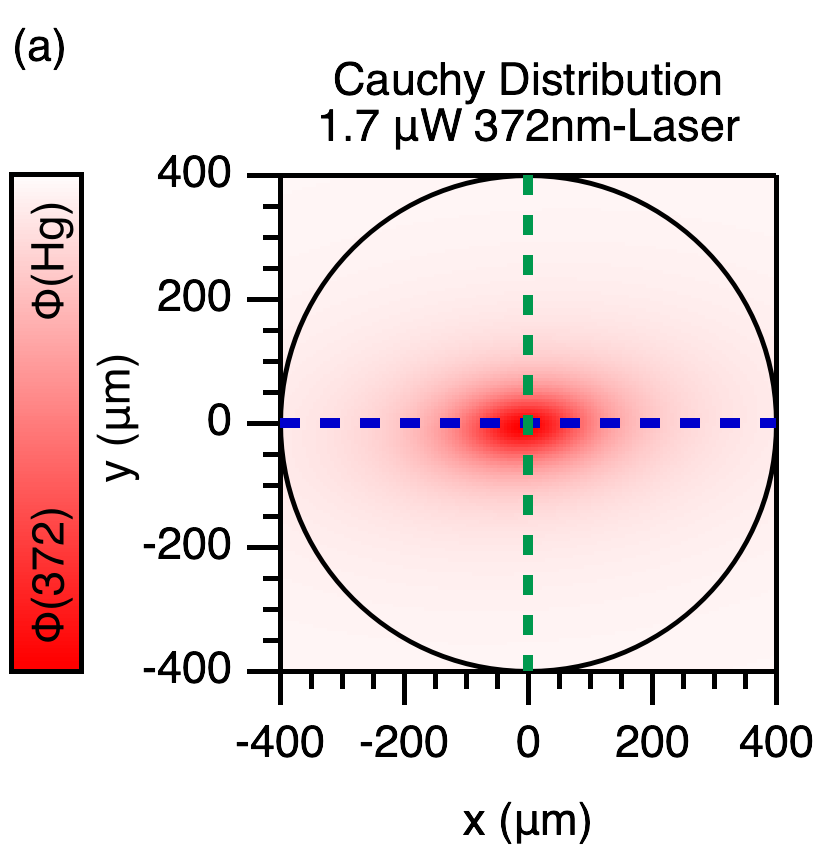}
        \includegraphics{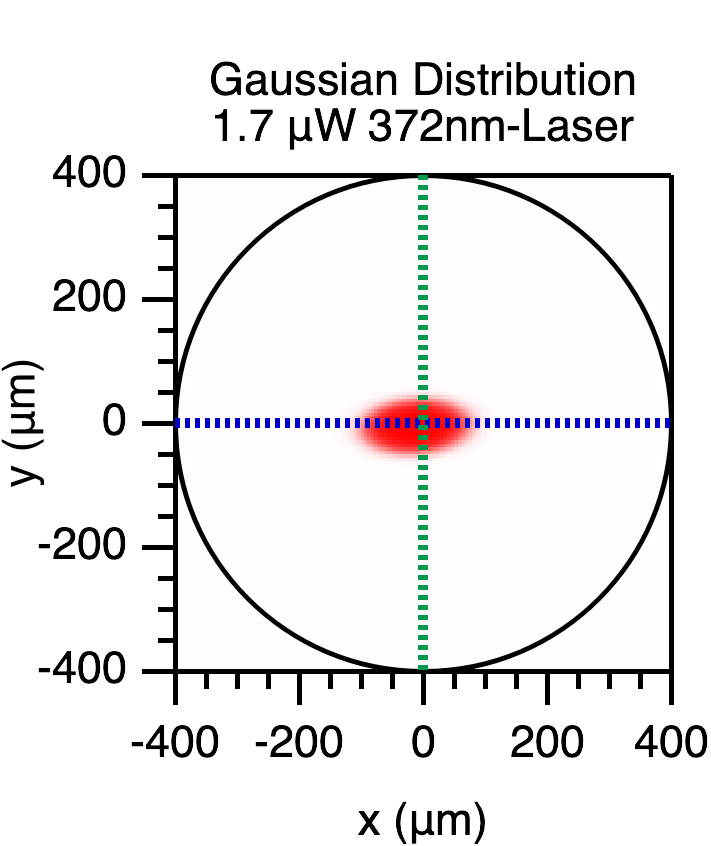}
        \includegraphics{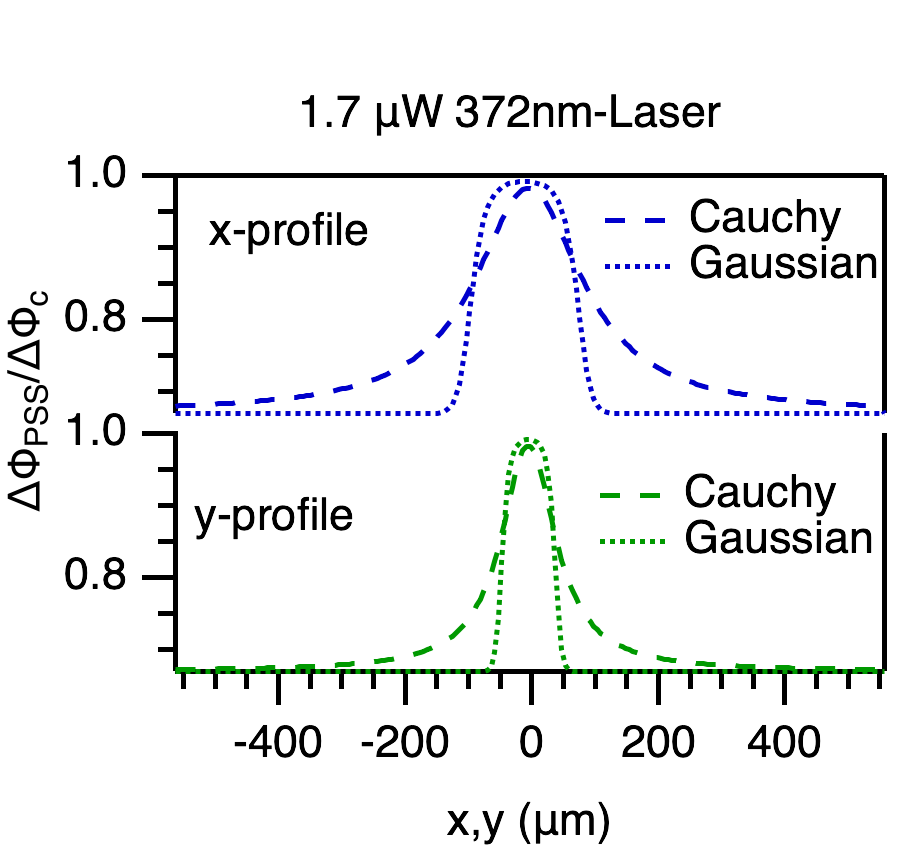}
        \includegraphics{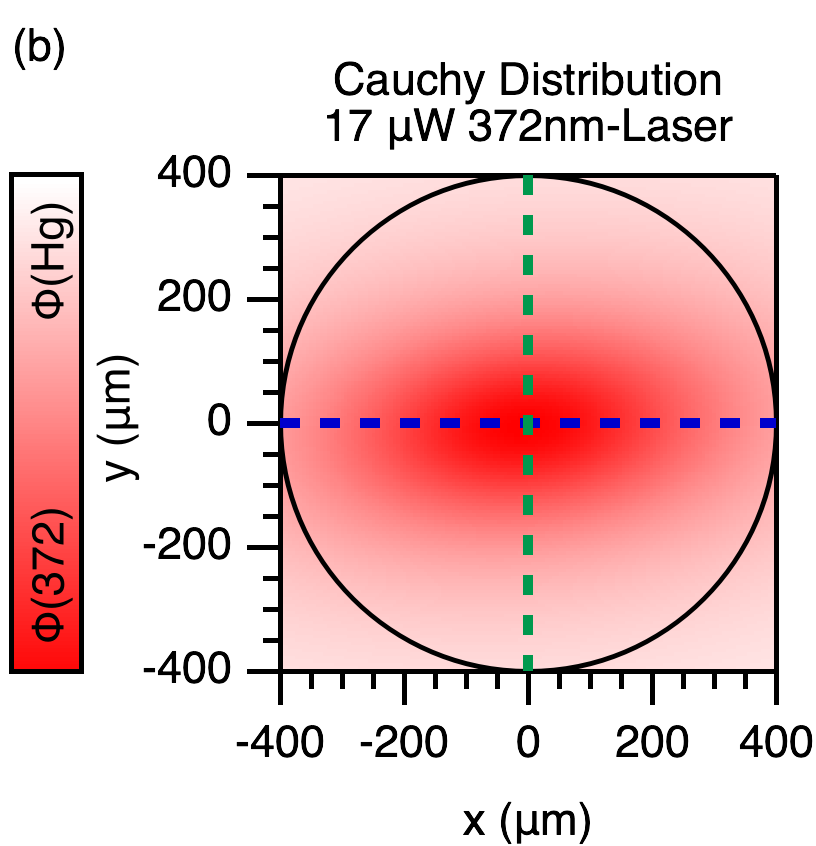}
        \includegraphics{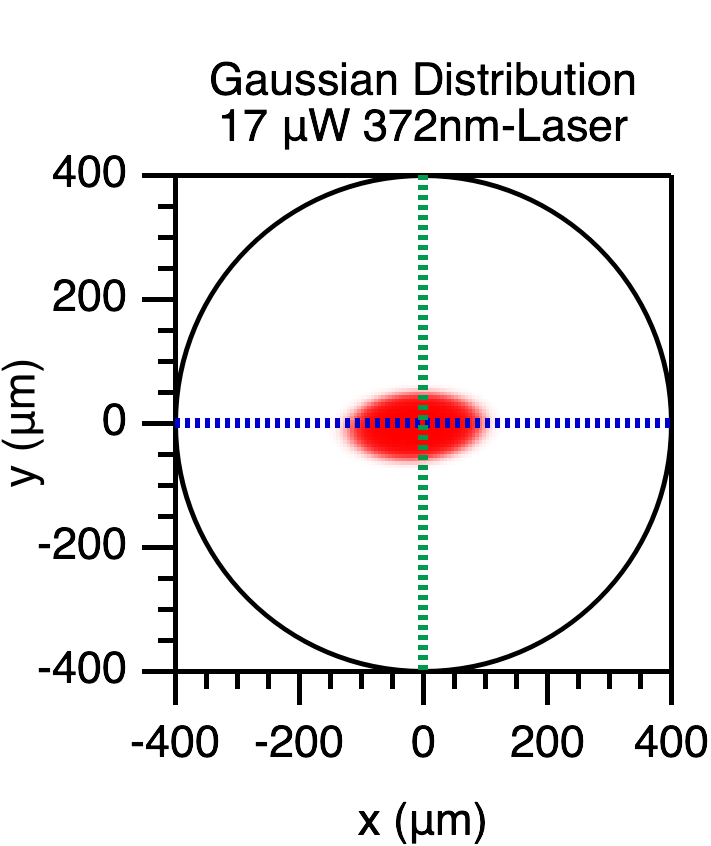}
        \includegraphics{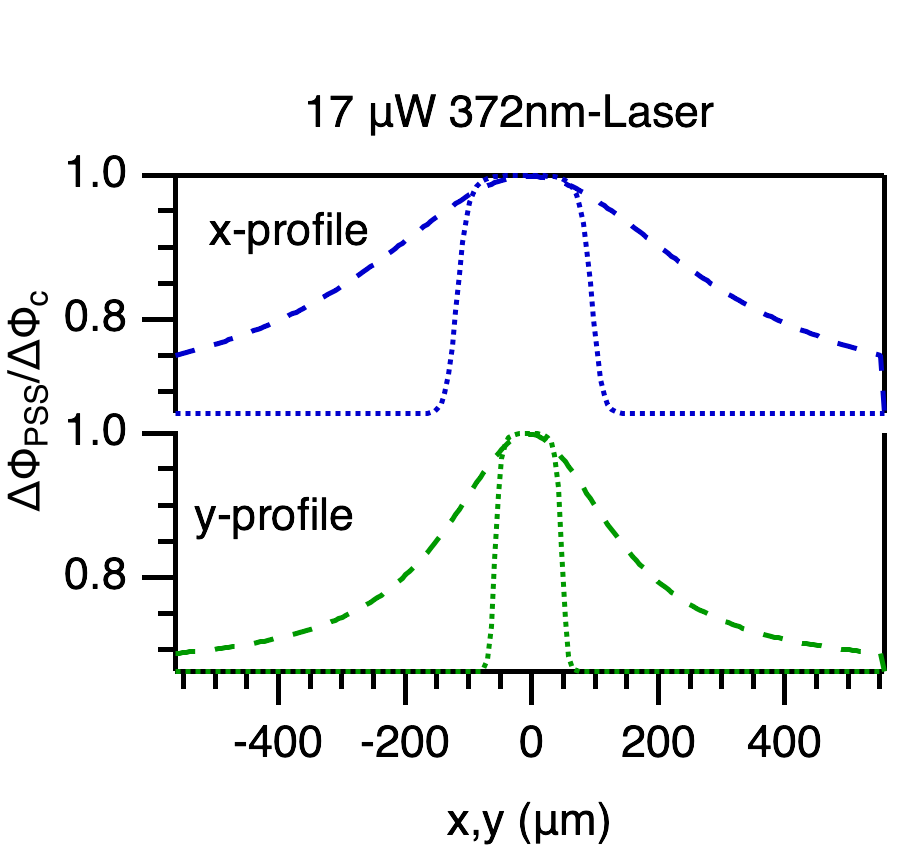}
        \caption{
        (a,b) Modelling of the work-function change based on 2D fits of the \qty{372}{nm} laser beam-profile using Cauchy and Gaussian distributions (left and middle panels). The modelling of the PEEM data was performed based on Eq.\,\ref{eq:phi_r} applied for every pixel for P$_L$= 1.7 (a) and \qty{17}{\micro W} (b), respectively. In the left panels we compare the vertical and horizontal profiles for Cauchy (dashed) and Gaussian (dotted) distributions. Note that changes in LEY and work function are proportional (see SM, Sec.~\textrm{III}).}
        \label{fig:PEEM_model}
\end{figure*}

\begin{figure}[h!]
        \includegraphics[width=8.6cm]{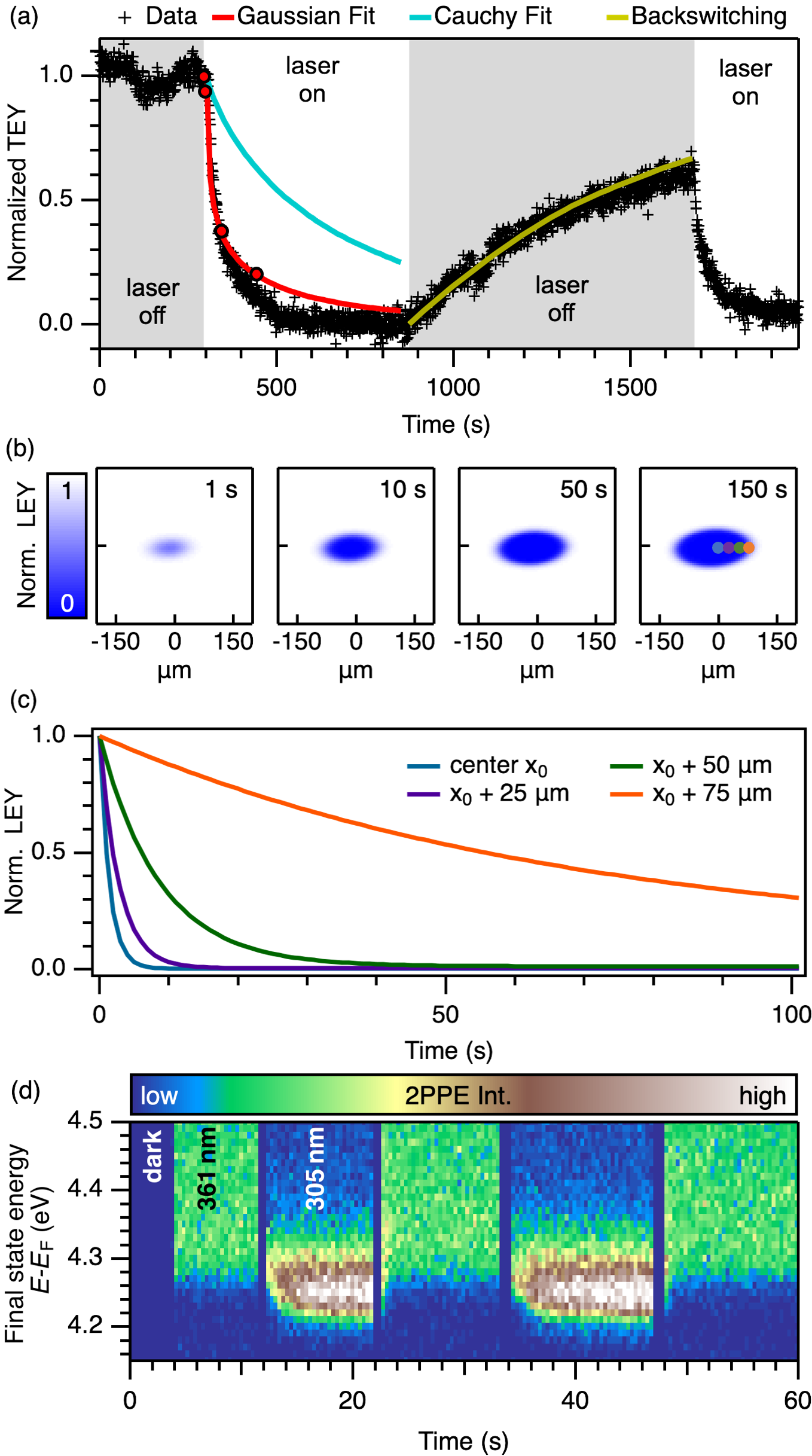}
          \caption{(a) Temporal evolution of the total electron yield (TEY) integrated over a FoV with a diameter of \qty{800}{\micro\meter} after turning on and off the cw laser with \qty{48}{\micro W} incident power. 
          We compare the acquired data (black markers) with the modelled curves for Gaussian (red curve) and Cauchy (cyan curve) shaped beam profiles. The yellow-green curve shows the expected behavior for switching back after turning off the laser.
        (b) Modelled spatial distribution of the LEY across the sample at \qtylist{1;10;50;150}{s} (cf.~red markers in panel a) after having turned on the laser with \qty{48}{\micro W} incident power. Norm. LEY=1 corresponds to the LEY at Hg-PSS and 0 to the LEY at the 372\,nm-PSS.
        (c) Local switching kinetics curves colored corresponding to the markers in panel b indicating their spatial positions: central position x$_0$ of the spot profile (blue), \qty{25}{\micro\meter} (violet), \qty{50}{\micro\meter} (green) and \qty{75}{\micro\meter} (orange) away from the spot center.
        (d) Single wavelength 2PPE measurements as a function of illumination time resolve the switching kinetics directly monitoring the work-function change. 
        }
\label{fig:kinetics}
\end{figure}

To spatially resolve the work-function shift across the sample, we recorded local electron yield (LEY) images in PEEM mode with and without laser illumination. 
The \qty{372}{nm} beam was focused in the upper left part of the FoV and impinged at a grazing angle of $22^{\circ}$ onto the sample. 
This expands the horizontal spot profile to a FWHM of around $67\,\unit{\micro\meter}$. 
Figures~\ref{fig:PEEM}b and c show the detector sensitivity corrected normalized images (LEY({\rm{Hg + 372}}) - LEY({\rm{Hg}}))/LEY({\rm{Hg}})=($\Delta$LEY/LEY$_0$) for identical spot profiles, but changing laser power by one order of magnitude.
Before image acquisition the sample was exposed to laser illumination for about 300\,s while the acquisition time for each images was about 150\,s.
The PEEM image in Fig.~\ref{fig:PEEM}b was recorded with a laser power of $1.7 \pm 0.3\,\unit{\micro\W}$. 
This corresponds to a photon flux of $\sim 1 \cdot 10^{17}\,{\rm{cm}}^{-2}{\rm{s}}^{-1}$ in the laser spot center, leading to about 1 switching event every \qty{10}{s}.\cite{moldt2015,bronsch2017a} 
At this threshold fluence we start to observe a local shift of the PSS state towards higher work function (higher density of \textit{cis}-isomers). 
From the horizontal and vertical profiles we estimate a spot size of $60 \times \qty{200}{\micro\meter}$, about $3-4$ times larger than the FWHM of the laser focus. 

Increasing the \qty{372}{nm} photon flux by a factor of 10, the area in Fig.~\ref{fig:PEEM}c showing a decreased EY is more pronounced and significantly larger. Horizontal and vertical profiles reveal a magnification of the projected laser focus FWHM by a factor of  $10$. While the differential photoemission intensity $\Delta$LEY changes gradually along the direction of laser incidence $x$, comparably sharp edges and a nearly flat top are observed along the $y$-direction.

Taking into account the laser-focus profile, we modeled the expected local work-function shifts $\Delta\Phi$ assuming first-order kinetics according to:
\begin{equation}
\Delta\Phi(\tilde{j})=\Delta\Phi_{\rm PSS}\frac{\tilde{r}}{\tilde{j}+\tilde{r}},
\label{eq:phi_r}
\end{equation}
For a detailed derivation of this equation, we refer to Ref.\,\onlinecite{bronsch2017a}. 
Here, $\tilde{j}=j({\rm 436\,nm})/j({\rm 372\,nm})$ corresponds to the photon flux ratio of Hg lamp and \qty{372}{nm} laser, while $\tilde{r}=\tilde{\sigma}({\rm 372\,nm})/\tilde{\sigma}({\rm 436\,nm})$ corresponds to the ratio of effective isomerization cross-sections for both light sources. 
$\Delta\Phi_{\rm PSS}$ denotes the work-function difference between the PSSs reached under 372 and 436\,nm illumination only. 
From the wavelength dependence of the effective isomerization cross-section across the S$_2$ absorption band we obtain $\tilde{\sigma}({\rm 372\,nm})\approx\tilde{\sigma}({\rm 365\pm 5\,nm})=1.4\cdot10^{-18}$\,cm$^2$.\cite{bronsch2017a,moldt2015} 
In the case of the S$_1$ band we approximate $\tilde{\sigma}({\rm 436\,nm})$ by $\tilde{\sigma}({\rm 455\,nm})=1.2\cdot10^{-18}$\,cm$^2$, which is known from \mbox{NEXAFS} experiments on Az11 SAMs.\cite{moldt2015}
For our simulations, we assumed homogeneous illumination by the Hg lamp in the PEEM's FoV and fitted the measured profile of the laser spot with either Cauchy or Gaussian distributions (see SM, Sec.\,\textrm{I}). 
Elliptical profiles account for the different widths in horizontal and vertical directions.
Figures~\ref{fig:PEEM_model}a and b show calculated local work-function changes for laser intensities of 1.7 and \qty{17}{\micro W}, respectively. The color scale ranges from the PSS of the Hg lamp ($\Phi({\rm{Hg}})$, white) to the maximal work-function change reached upon \qty{372}{nm} illumination only ($\Phi({\rm{372}})$, red), \textit{i.e.} the PSS with predominantly \textit{cis} Az11 chromophores. 
As expected, the Cauchy distribution gives smooth changes of the work function over several hundreds of micrometer away from the spot center at $(0,0)$. 
In contrast, for a Gaussian distribution we observe work-function profiles with steepened edges and a flat top. 
None of the models perfectly describes the measured laser spot, but a combination of both contains all characteristics of the switched area. 
The shape of the beam profile is thus essential for the work-function gradient from the center of the switched spot to the non-changed environment, whereas the local photon flux ratio of the laser beam and the Hg lamp determines the magnitude of the local work-function shift. 
The tails of the Cauchy profile $\propto 1/x^2$ lead to an extended switched area, those of a Gaussian $\propto \exp(-x^2)$ to steepened edges and a pronounced flat top along the $y$-direction. 
An increase of the local work function $\Delta\Phi$ upon \textit{trans}-\textit{cis} isomerization leads to a decrease of the local electron yield $\Delta$LEY. 
In Sec.~\textrm{III} of the SM we demonstrate $\Delta\Phi \propto -\Delta$LEY. 
Thus simulated patterns of $\Delta\Phi$ and measured $\Delta$LEY profiles are quantitatively comparable.
Comparing the simulations (cf. Fig. \ref{fig:PEEM_model}) with our data (cf. Fig. \ref{fig:PEEM}), we conclude that only those areas hit by the laser beam are switched. In the PEEM experiment we spatially resolve the photoelectron yield and not directly the work function. 

The photon flux in the laser spot not only varies the local PSS, but also the kinetics to reach this PSS. As a measure for the overall switching kinetics, we evaluate the transient of the total electron yield TEY($t$) by summing up the electron count-rate in the complete FoV as a function of exposure time $t$. 
As shown in Fig.~\ref{fig:kinetics}a, TEY($t$) decreases as soon as the laser beam illuminates the sample, since the Az11 SAM changes towards the new PSS that includes overall a higher fraction of \textit{cis} isomers.\cite{ahqune2008,bronsch2017,bronsch2017a}  
When turning laser illumination off, the ensemble of Az11 molecules relaxes back to the Hg-PSS. 

To describe the transient work-function shift $\Delta\Phi(t)$, which expresses the ensemble switching kinetics, the time-dependent change of the local fraction of \textit{cis} molecules $\Delta\chi_{\rm c}(t)$ has to be considered (for details see SM, Sec.~\textrm{IV}):
\begin{eqnarray}
\Delta\Phi(x,y,t) & = & \Delta\Phi_{\rm PSS} \cdot \Delta\chi_{\rm c}(x,y,t)
\end{eqnarray}
with
\begin{eqnarray}
\Delta\chi_{\rm c}(x,y,t) & = & \chi_{\rm c}(x,y,t) - \chi_{\rm c}(0) \nonumber \\
& = & \Delta\chi_{\rm c,PSS}(x,y) \cdot
( 1- e^{-t\cdot(j_1(x,y) \cdot \tilde{\sigma}_{1} + j_2 \cdot \tilde{\sigma}_{2})}),\nonumber
\end{eqnarray}
with $\Delta\chi_{\rm c,PSS}(x,y)=\chi_{\rm c}(x,y,\infty) - \chi_{\rm c}(0)$, $j_{1,2} =  j({\rm 372\,nm}),\,j({\rm 436\,nm})$ and $\tilde{\sigma}_{1,2} = \tilde{\sigma}({\rm 372\,nm}),\,\tilde{\sigma} ({\rm 436\,nm})$. 
\\ 

The solid lines in Fig.~\ref{fig:kinetics}a show the calculated change of the TEY for the fitted Cauchy and Gaussian laser spot profiles (cf.~Fig.~\ref{fig:PEEM}). 
Using the Cauchy distribution, we get poor agreement between modelled and measured switching kinetics (cyan line). 
The Gaussian beam profile (red line) shows a much better agreement. 
This matches the observation of sharp edges of the switched areas, shown by the line profiles in Fig.~\ref{fig:PEEM}c.
We modelled the temporal evolution of the spatial work-function distribution in a defined FoV to illustrate how the LEY gradually evolves under laser illumination. 
Figure~\ref{fig:kinetics}b shows exemplary images after \qtylist{1;10;50;150}{s} laser illumination with a Gaussian profile, Fig.~\ref{fig:kinetics}c the corresponding kinetics curves for selected positions in real space.
In the center of the beam, the PSS state is close to a pure \textit{cis} SAM and is reached after few seconds for \qty{48}{\micro W} laser power (blue curve). 
In the tails of the laser spot, the local PSS is not reached within \qty{100}{s} (orange curve).  
The fast and precise work-function tuning modelled for the spot center can be shown using photoemission from a pulsed laser source operated at a wavelength in the S$_2$ band. 
This gives excitation densities high enough to induce two-photon photoemission (2PPE) and enables direct tracing of the work-function change as shown in Fig.~\ref{fig:kinetics}d for \qty{361}{} and \qty{305}{nm} photons. 
Upon excitation with \qty{305}{nm}, the low energy cutoff of the photoemitted electron spectrum shifts towards \qty{4.22}{eV} with respect to \qty{4.27}{eV} when illuminated by \qty{361}{nm}. 
This indicates a work-function shift of \qty{50}{meV}. 
Enhancing the capabilities of the experimental PEEM setup with a new pulsed laser source (SM, Sec.~\textrm{V}) will enable spatially resolved measurements of the work-function distribution, paving the path for transient \si{\micro\meter} work-function patterning beyond the LEY signatures.

In conclusion, we showed that well-defined and homogeneous switched areas can be produced by illuminating an azobenzene-functionalized surface with a focused UV beam without the need of current patterning techniques with predefined vignettes\cite{yue2020,bakker2023}, enabling on-the-fly modifications. 
When shaping the beam profile, the absolute value of the local work function as well as the time scale required to switch the work function can be controlled. The switched area and isomerization profile depend on the laser spot profile paving the route towards designing cheap organic optoelectronic devices with active area, tuning not only the work function, but, e.g., the lateral excitonic coupling among the photoswitches.\\ 

We thank Rafal Klajn for the kind supply of the Az11 molecules and Johannes Mosig for fruitful discussions and computational support. J.B.~thanks the International Max Planck Research School for Elementary Processes in Physical Chemistry for financial support in the initial phase of his PhD. This project was supported by Deutsche Forschungsgemeinschaft via the Collaborative Research Center TRR 227 on Ultrafast Spin Dynamics, projects A01 and B07. Part of the equipment was financed by the Deutsche Forschungsgemeinschaft via the program for major research instrumentation (91b GG).

\end{document}


\title{Supplementary Material: Local work-function manipulation by external optical stimulation}
\author{Jan B\"ohnke}
\affiliation{Freie Universit{\"a}t Berlin, Fachbereich Physik, Arnimallee 14, 14195 Berlin, Germany}
\author{Beatrice Andres}
\affiliation{Freie Universit{\"a}t Berlin, Fachbereich Physik, Arnimallee 14, 14195 Berlin, Germany}
\author{Larissa Boie}
\affiliation{Freie Universit{\"a}t Berlin, Fachbereich Physik, Arnimallee 14, 14195 Berlin, Germany}
\author{Angela Richter}
\affiliation{Freie Universit{\"a}t Berlin, Fachbereich Physik, Arnimallee 14, 14195 Berlin, Germany}
\author{Cornelius Gahl}
\affiliation{Freie Universit{\"a}t Berlin, Fachbereich Physik, Arnimallee 14, 14195 Berlin, Germany}
\author{Martin Weinelt}
\email[Corresponding author: ] {weinelt@physik.fu-berlin.de}
\affiliation{Freie Universit{\"a}t Berlin, Fachbereich Physik, Arnimallee 14, 14195 Berlin, Germany}
\author{Wibke Bronsch}
\email[Corresponding author: ] {wibke.bronsch@fu-berlin.de}
\altaffiliation[Current address: ]{Elettra - Sincrotrone Trieste S.C.p.A., Strada Statale 14 - km 163.5 in AREA Science Park, 34149 Basovizza, Trieste, Italy}
\affiliation{Freie Universit{\"a}t Berlin, Fachbereich Physik, Arnimallee 14,
14195 Berlin, Germany}
\date{\today}

\maketitle
\tableofcontents
\section{2D fitting of the spot profile}
For the spatially resolved modelling of the work-function changes shown in Fig.~3 of the main text, we were using the fit results for the measured 372-nm laser spot-profile based on elliptical Cauchy and Gaussian distributions:

\begin{align}
\label{eq:fitfunctions}
f_{\rm Cauchy}(x,y)&=\frac{w}{\pi (w^2+a^2 x^2 + b^2 y^2)},\nonumber\\
f_{\rm Gaussian}(x,y)&=\frac{e^{-\left(\frac{a^2 x^2+b^2 y^2}{2 w^2}\right)}}{\sqrt{2\pi}w}.
\end{align}\\

\begin{figure*}[t]
        \includegraphics{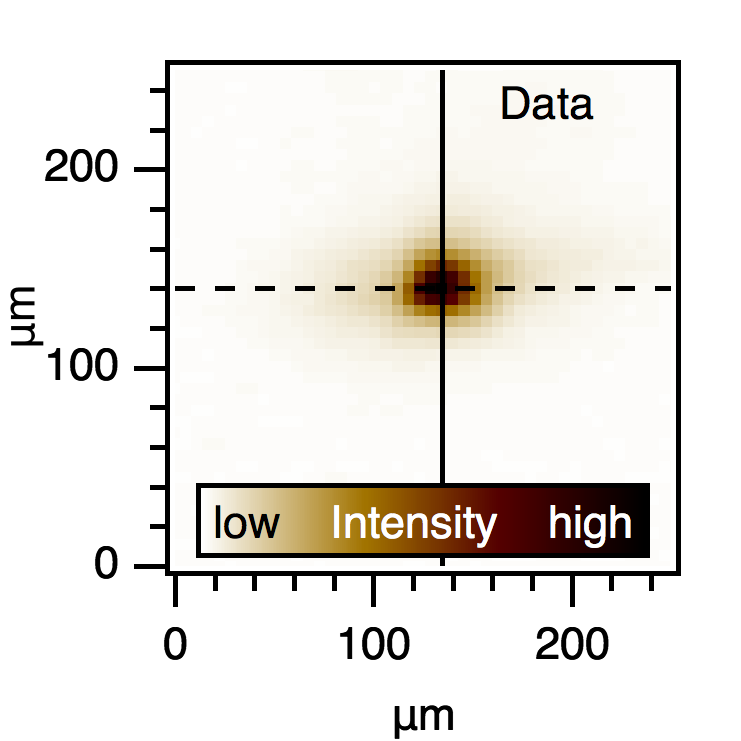}
        \includegraphics{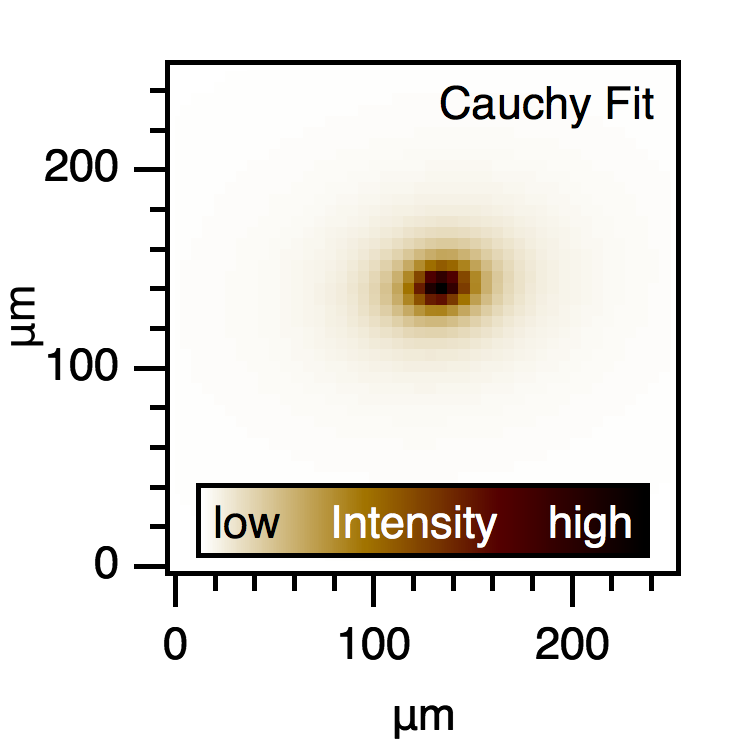}
        \includegraphics{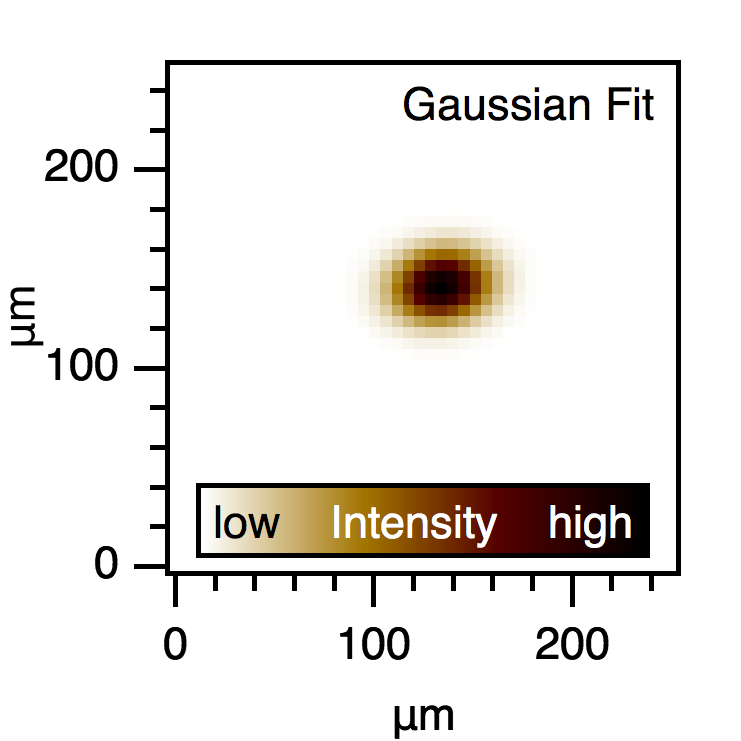}
        \includegraphics{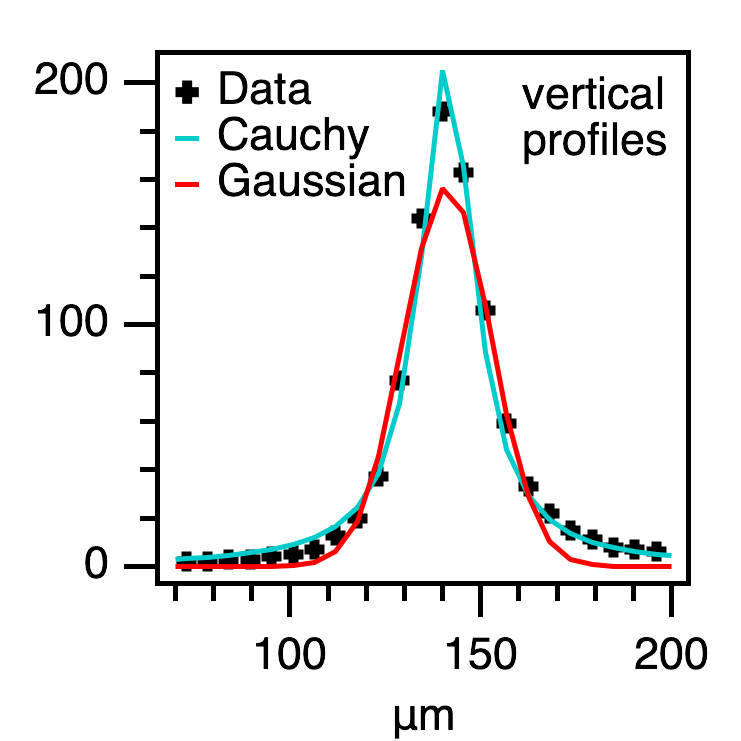}
        \includegraphics{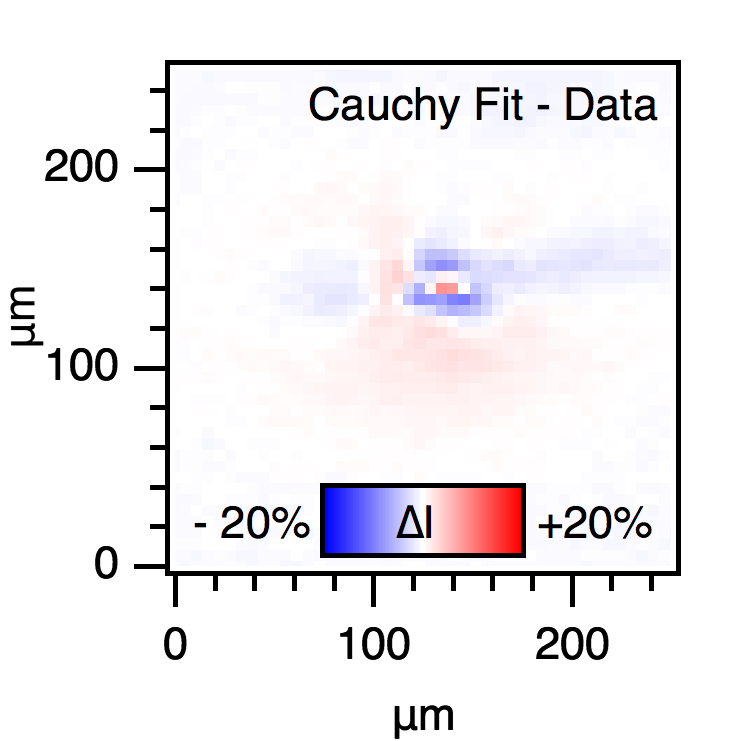}
        \includegraphics{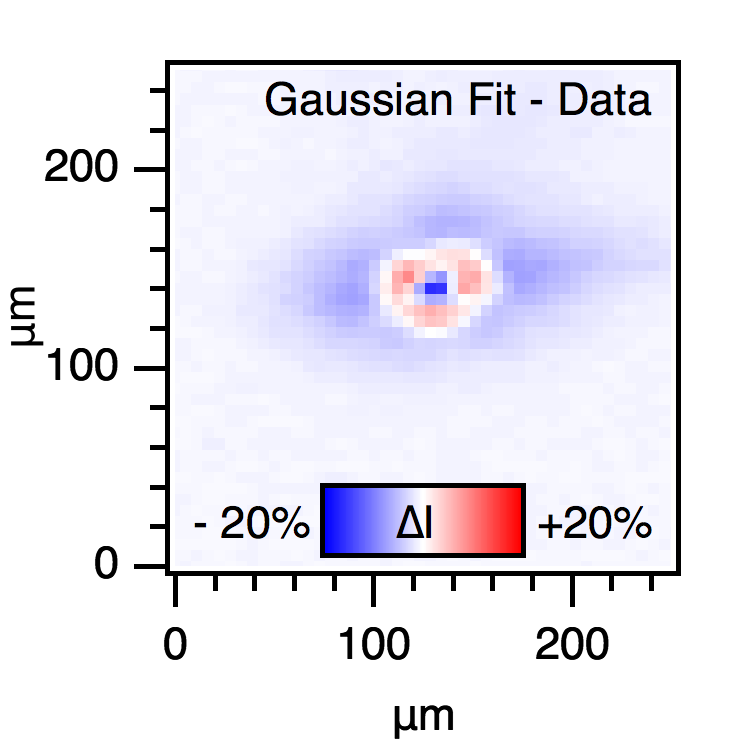}
        \caption{2D fits of the 372-nm laser spot-profile. The measured profile (upper left panel) was modelled with an elliptical Cauchy distribution (upper middle panel) and an elliptical Gaussian distribution (upper right panel). The fit quality is indicated in the lower panels, comparing an exemplary cut through the measured beam profile and difference maps between fits and data.
       }
        \label{fig:SOM_2D_fits}
\end{figure*}

Figures~\ref{fig:SOM_2D_fits} compare the 2D fits to the measured spot profile recorded on a CCD camera placed in the focal plane for normal incidence. Fit parameters are listed in Tab.\,\ref{tab:fitparameters}. Since the laser beam hits the sample surface under an angle of incidence of 22$^{\circ}$, fitted spot profiles used for the simulations in Fig.~3 of the main text were stretched in $x$-direction accordingly by a factor of 2.67.

\begin{table}[h!]
\label{tab:fitparameters}
\begin{tabular}{p{3cm}|p{3cm}|p{3cm}|p{3cm}}
 & w & a &  b\\\hline
Cauchy Fit&7.69$\pm$0.05&5.23$\pm$0.03&3.80$\pm$0.02\\\hline
Gaussian Fit&7.86$\pm$0.05&3.85$\pm$0.03&2.88$\pm$0.02\\
\end{tabular}
\caption{Parameters of the 2D fits to the laser spot according to Eqs.\,\ref{eq:fitfunctions}. Fits were performed per pixel. Pixel size is 5.6\,\unit{\micro\meter} $\times$ 5.6\,\unit{\micro\meter}.}
\end{table}

\section{Photostationary state induced by the mercury lamp}
\begin{figure*}[t]
        \includegraphics{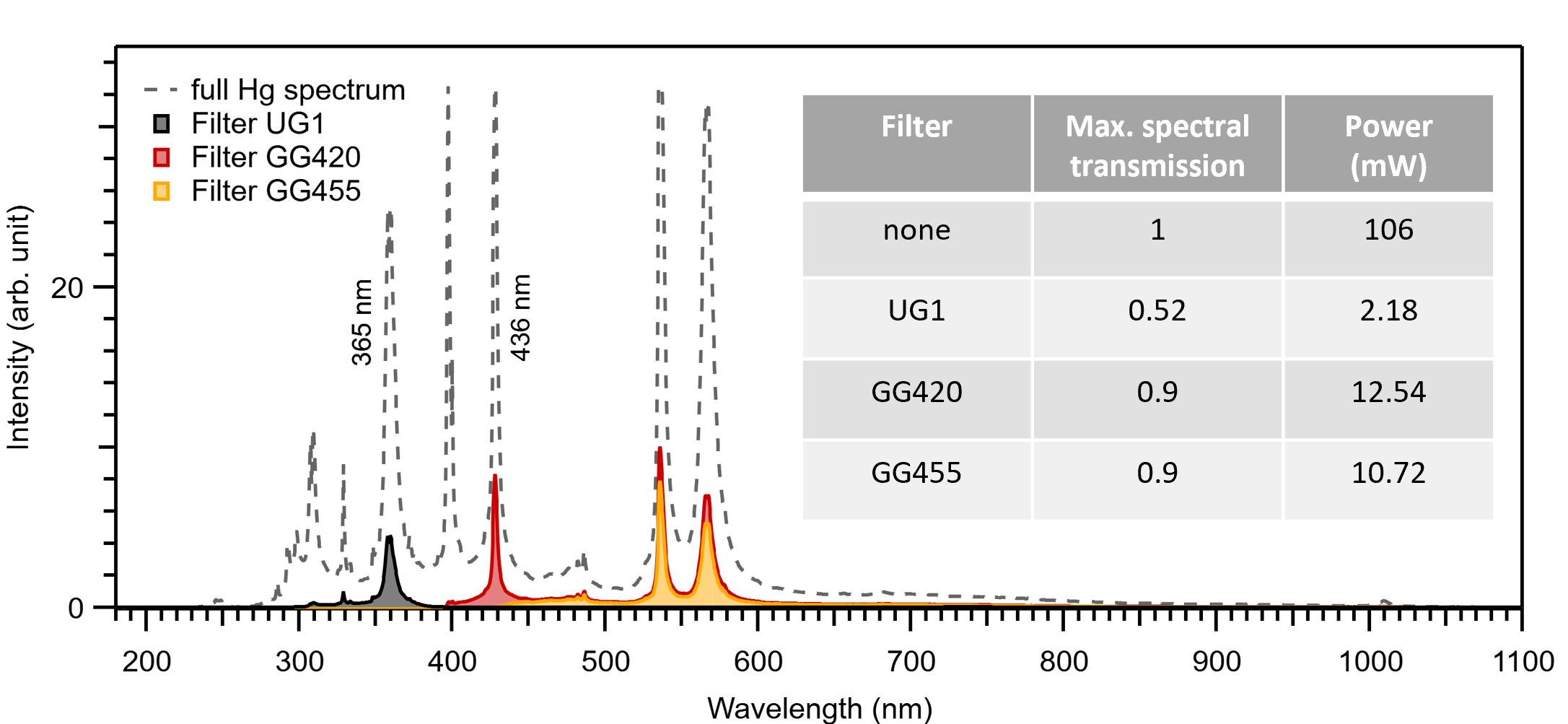}
        \caption{Hg-lamp emission spectrum (dashed grey) and partial contributions after applying different optical filters. The inset table shows the Hg-lamp power measured after the different optical filters used to calculate the photon flux at 436 and 365\,nm.}
        \label{fig:Hg_spectra}
\end{figure*}

During the experiments we illuminated the samples with a mercury short-arc lamp (Osram HBO \qty{100}{\watt/2}) to photoemit electrons from the azobenzene-functionalized gold surface. 
As discussed, the spectrum of the Hg lamp has several emission lines among which two main lines induce switching of the azobenzene molecules inbetween {\it cis} and {\it trans} configurations. 
In order to determine the photostationary state reached by illumination with the Hg lamp (Hg-PSS) we analyzed the emission spectrum (recorded by an AvaSpec-ULS4096CL-RS-EVO spectrometer) and extracted the relative contribution of the two emission lines at 365 and \qty{436}{nm} to the total emitted power. Figure~\ref{fig:Hg_spectra} shows the full spectrum of the Hg lamp (dashed line) and the remaining contributions after inserting different optical filters (black, red, and yellow areas). 
We estimated the flux $j_{\rm 436\,nm}$ using the difference in measured power $P$ (Melles Griot 13 PEM 001/J) between filters GG420 and GG455, which select the photons emitted at 436\,nm. 
We considered uniform illumination of a unit area of $A=\pi\cdot \qty{1}{cm\squared}$ and an illumination interval of $\Delta t= \qty{1}{s}$.
During the actual experiments the Hg lamp was operating at the total power of $P_{\rm exp}=144$\,mW. This corresponds to a photon flux of 
\begin{align*}
j_{\rm 436\,nm} =\frac{\frac{P_{\rm GG420}/T_{\rm GG420} - P_{\rm GG455}/T_{\rm GG455}}{P_{\rm none}}\cdot P_{\rm exp}}{E_{\rm ph}({\rm 436 nm)}\cdot A} 
\approx 1.9\cdot 10^{15} \frac{1}{{\rm cm}^2\,{\rm s}}.
\end{align*}
The photon flux of the \qty{365}{nm} emission line corresponds to light passing filter UG1
\begin{align*}
j_{\rm365\,nm} =\frac{\frac{P_{\rm UG1}/T_{\rm UG1}}{P_{\rm none}}\cdot P_{\rm exp}}{E_{\rm ph}({\rm 365 nm)}\cdot A}
\approx 3.3\cdot 10^{15}\frac{1}{{\rm cm}^2\,{\rm s}}.
\end{align*}\\

We assume that the isomerization cross-section is constant across the S$_1$ absorption band and approximate $\tilde{\sigma}_{\rm iso}(\qty{436}{nm})\approx \tilde{\sigma}_{\rm iso}(\qty{455}{nm})=\qty{1.2e-18}{cm\squared}$ [\onlinecite{moldt2016}]. Furthermore, using $\tilde{\sigma}_{\rm iso}(\qty{365}{nm})=\qty{1.4e-18}{cm\squared}$ [\onlinecite{moldt2016}],
we calculated the fraction of {\it cis} molecules in the Hg-PSS to [\onlinecite{Bronsch2017a}]:

\begin{equation*}
\chi_{\rm c}=\frac{{\sigma_{\rm 365\,nm}}/{\sigma_{\rm 455\,nm}}}{{j_{\rm 436\,nm}}/{j_{\rm 365\,nm}}+{\sigma_{\rm 365\,nm}}/{\sigma_{\rm 455\,nm}}} \approx 67 \%.
\end{equation*}\\

\begin{figure*}[h!]
        \includegraphics{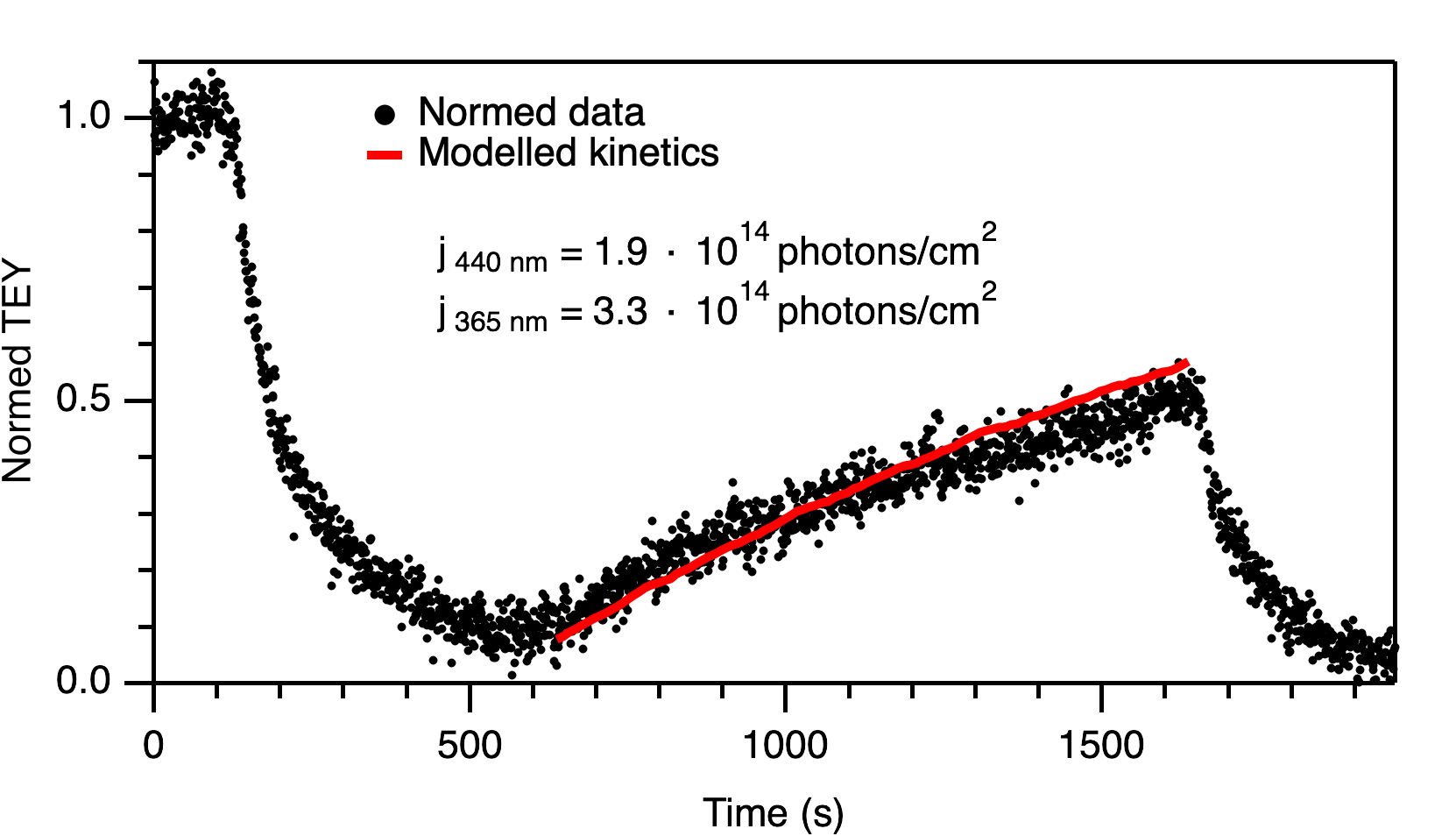}
        \caption{Back-switching kinetics for pure Hg-lamp illumination into the related PSS. The net photon fluxes of the 365 and 440\,nm spectral lines of the Hg lamp amount to $j_{\rm440\,nm}=1.9\cdot10^{14}\,{\rm cm}^{-2}$ and $j_{\rm365\,nm}=3.3\cdot10^{14}\,{\rm cm}^{-2}$ in order to reach a reasonable description of the measured kinetics (red solid line). Deviations are attributed to fluctuations of the Hg-lamp.
       }
        \label{fig_SOM:Hg_kinetics}
\end{figure*}

To prevent beam damage of the SAM the flux of the Hg-lamp was further reduced using grey filters. We estimated the number of photons arriving at the sample from the back-switching kinetics after laser illumination. The work-function shift is proportional to the change in total electron yield $\Delta\Phi  \propto -\Delta$TEY, as demonstrated in Section~{\textrm{III}}. Figure~\ref{fig_SOM:Hg_kinetics} shows the transient evolution of the TEY for the measurement conditions presented in Fig.~3 of the main text. We varied the photon flux with constant ratio of $j_{\rm365\,nm}/j_{\rm 436\,nm} = 3.3/1.9$ to model the back-switching kinetics to the Hg-PSS (red solid line in Fig.~\ref{fig_SOM:Hg_kinetics}).

\clearpage
\section{Conversion of sample work-function into photoelectron yield}
\begin{figure*}[t]
        \includegraphics{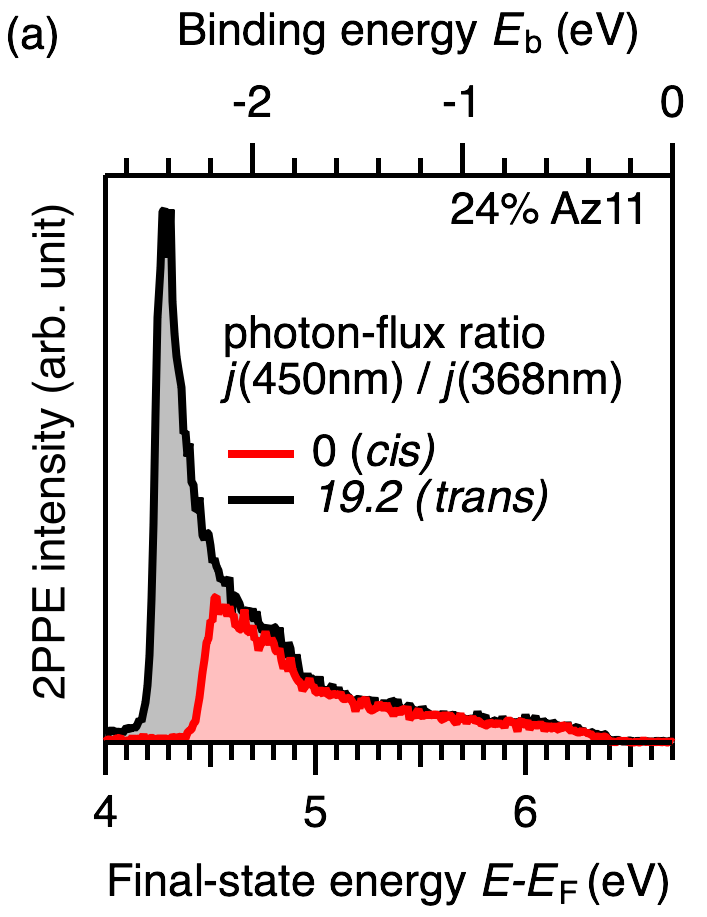}   
   \includegraphics{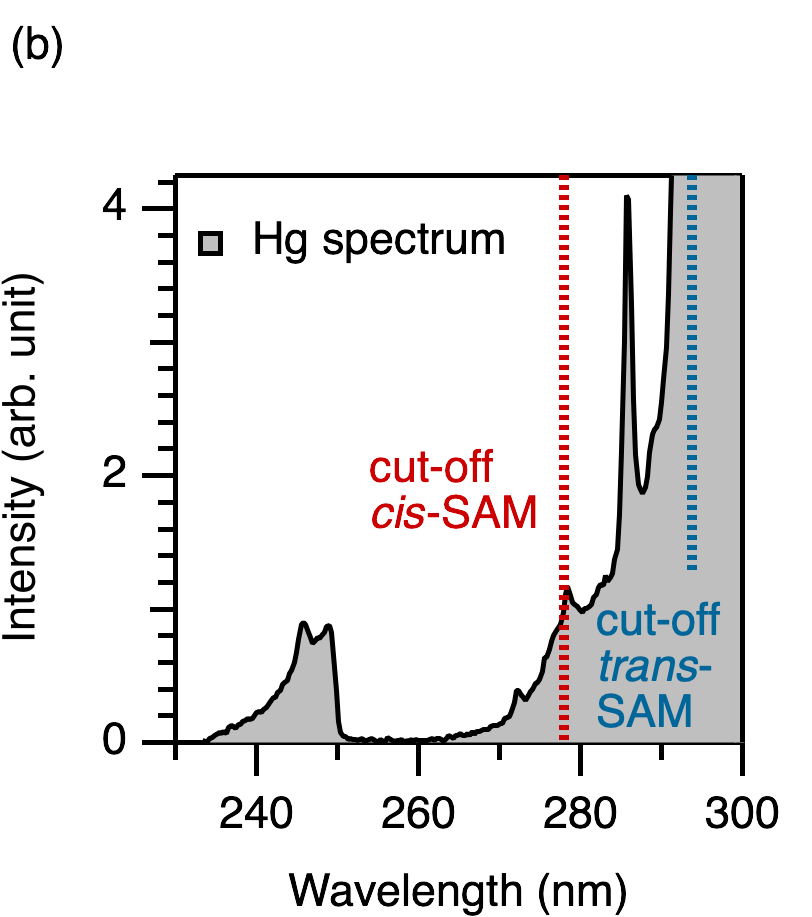}
   \includegraphics{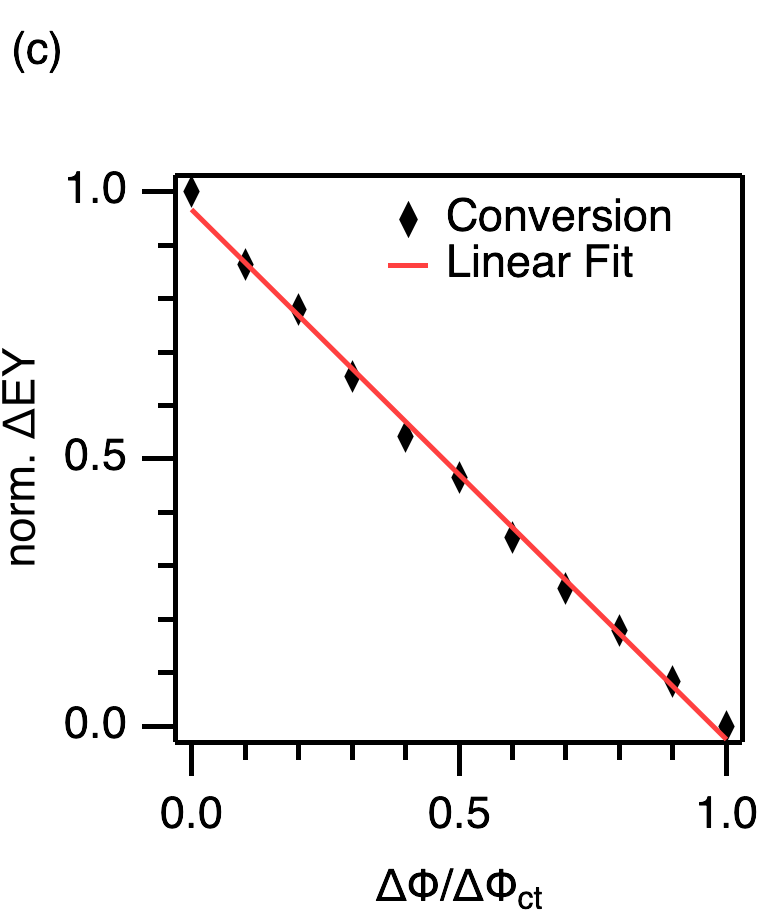}
         \caption{(a) Energy-resolved two-photon photoemission spectra recorded for a 24\,\% Az11 SAM in {\it trans} and {\it cis} configuration. (b) Section of the emission spectrum of the Hg lamp. The dashed red and blue vertical lines mark the cut-off wavelengths of photons leading to photoelectron emission from {\it cis} and {\it trans} SAMs, respectively. (c) Change in the photoelectron yield as a function of the change in sample work-function when switching from {\it trans} to {\it cis} SAM. We apply Eq.\,\ref{eq:LEY} taking into account the energy distribution curves (EDCs) in panel (a) and the spectral distribution $I(h\nu)$ of the Hg lamp in panel (b).
       }
        \label{fig:EYconversion}
\end{figure*}

In Ref.~\onlinecite{Bronsch2017} we reported a series of energy distribution curves (EDCs) of Az11 samples measured by two-photon photoemission (2PPE) spectroscopy with a chromophore density diluted to 24\,\% of a densely packed SAM.
We were probing the sample with a pulsed 368-nm laser beam for switching into a pure {\it cis} PSS and inducing photoemission via two-photon absorption (red spectrum in Fig.~\ref{fig:EYconversion}a). 
Additional irradiation with a 450-nm cw-laser reduces the amount of {\it cis} molecules. Using a photon flux $j_{\rm 450\,nm}$ 19 times higher than $j_{\rm 368\,nm}$, we reached a PSS corresponding to a nearly pure {\it trans} SAM. The work function of the respective PSSs can be read out from the low energy cut-off at the bottom axis of Fig.~\ref{fig:EYconversion}a showing the final state energy. 
The top axis shows the binding-energy of the initial state referencing to the Fermi level $E_{\rm F} = 0$. 
The measured photoelectron yield stems mainly from the Au substrate and depends on the energy difference between the sample work-function and the photon energy. The Hg lamp has a range of photon energies which can induce photoemission.

Figure~\ref{fig:EYconversion}b shows a section of the emission-spectrum of the Hg-lamp in the relevant wavelength range. Vertical red and blue dashed lines mark the cut-off wavelength  for photoemission from a {\it cis} and {\it trans} SAM, respectively (the cut-off wavelength corresponds to the photon energy equal to the sample work-function).
As outlined in Sec.~\textrm{II}, the Hg-PSS has a fraction of \qty{67}{\percent} {\it cis} Az11 molecules. 
Its work function amounts to about \qty{4.35}{eV} (\qty{285}{nm} in Fig.~\ref{fig:EYconversion}b).\cite{Bronsch2017}

We describe the local electron yield LEY$(x,y)$ via the local work function $\Phi(x,y)$ according to: 
\begin{equation}
{\rm LEY} (\Phi(x,y)) =\int_{\Phi}^{h\nu_{\rm max}} \int_{-(h\nu-\Phi(x,y))}^{0} {\rm EDC}(E_{\rm b})I(h\nu){\rm d}E_{\rm b} {\rm d}h\nu.
\label{eq:LEY}
\end{equation}
Here, EDC($E_{\rm b}$) corresponds to the photoelectron spectrum shown in Fig.~\ref{fig:EYconversion}a and the light intensity $I$ as a function of photon energy $h\nu$ is calculated from the spectrum of the Hg lamp shown in Fig.~\ref{fig:EYconversion}b ($I(h\nu)d\nu = I(\lambda) (c/\lambda^2) d\lambda$).
The highest available photon energy $h\nu_{\rm max} =5.39$\,eV is determined by the spectral cut-off of the Hg lamp at 230\,nm.
Figure~\ref{fig:EYconversion}c demonstrates that the electron yield decreases linearly with the change in work function. Note that we consider only small work-function changes of about 30\,meV. The total electron yield (TEY) which is discussed in the main text is then given by the sum over all LEY$(x,y)$ in the field of view of the PEEM measurement. We normalized the TEY to describe the change between the HG-PSS state ($\text{TEY}=1$) and the PSS reached by additional cw-laser irradiation ($\text{TEY}=0$).

\section{Fraction of cis molecules during multi-wavelength irradiation}
The fraction of {\it cis} molecules $\chi_{\rm c}$ of the total number $N$ of Az11 molecules in a SAM reached for a certain PSS is defined by
\begin{equation}
\chi_{\rm c}(\lambda)=\frac{N_{\rm c}}{N}=\frac{j(\lambda)\cdot\sigma_{\rm isom, tc}(\lambda)}{j(\lambda)\cdot\sigma_{\rm isom, tc}(\lambda)+j(\lambda)\cdot\sigma_{\rm isom, ct}(\lambda)},
\end{equation}
considering the photon flux $j$ and the isomerization cross sections $\sigma_{\rm isom}$ for {\it trans-cis} and {\it cis-trans} isomerization.
Note that here we are considering exclusively optically induced switching processes, neglecting thermally induced back-switching from the metastable {\it cis} into the {\it trans} configuration.
The switching kinetics induced by irradiation of the sample with light of a wavelength in the range of the $S_1$ or $S_2$ absorption band of the chromophores is described by the transient evolution of the fraction of {\it cis} molecules:
\begin{equation}
\chi_{\rm c}(t,\lambda)=\chi_{\rm c}(\infty,\lambda)+\left(\chi_{\rm c}(0,\lambda)-\chi_{\rm c}(\infty,\lambda)\right)\cdot e^{-t\cdot j_{\lambda}\cdot\tilde{\sigma}_{\lambda}},
\end{equation}
with $\tilde{\sigma}_{\lambda}=\sigma_{\rm isom, tc}(\lambda)+\sigma_{\rm isom, ct}(\lambda)$.
When using more than one wavelength that induces switching of the molecules, we need to add up the contributions of each irradiation source, leading to
\begin{equation}
\chi_{\rm c}(t,\lambda_1,...,\lambda_{\rm n})=\chi_{\rm c}(\infty,\lambda_1,...,\lambda_{\rm n})+\left(\chi_{\rm c}(0,\lambda_1,...,\lambda_{\rm n})-\chi_{\rm c}(\infty,\lambda_1,...,\lambda_{\rm n})\right)\cdot e^{-t\cdot \sum_{i} j_{\lambda_{i}}\cdot\tilde{\sigma}_{\lambda_{i}}}.
\end{equation}
Note that every additional wavelength at which the molecule absorbs light will accelerate the switching kinetics, leading to a faster arrival in the new PSS.

\section{Characterization of the spatial resolution of the momentum microscope} \label{sec:charac_ToFMM}

\begin{figure*}[t]
       \includegraphics[width=17cm]{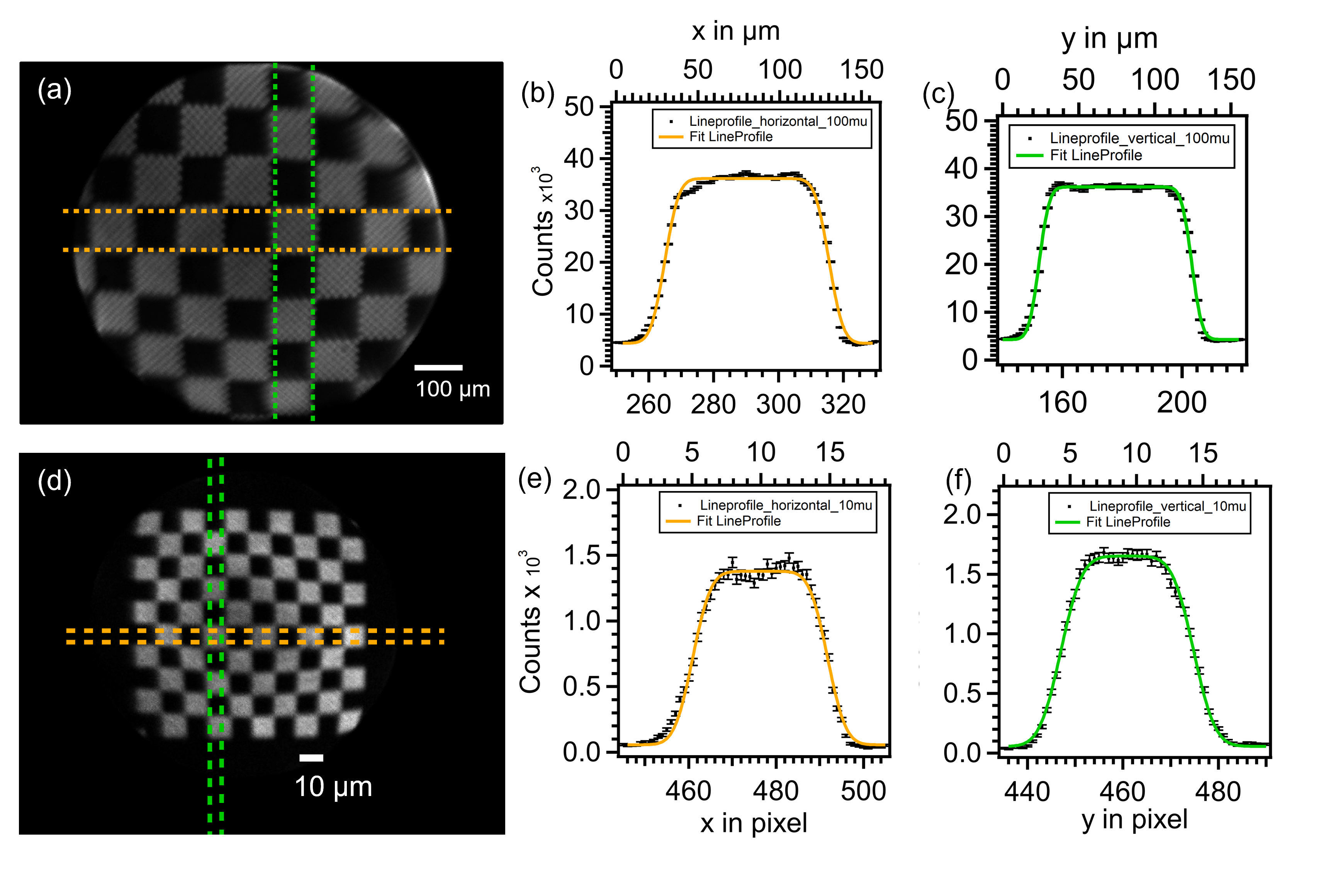}
        \caption{(a) Spatially-resolved image of the Chessy calibration sample showing a field of view of \qty{800}{\micro\meter} diameter. The same settings were used for the Az11 SAM PEEM images shown in the main text. Orange and green dashed lines indicate the line profile cuts which are presented in (b) and (c) with the corresponding Gaussian-broadened step-function fit, \textit{cf.}~Eq.~\ref{eq:error_func_fit}. (d) Spatial image of the Chessy calibration sample showing a field of view with a diameter of \qty{100}{\micro\meter} with the corresponding line profile cuts and fits in (e) horizontal (orange) and (f) vertical direction (green).}
        \label{fig:SOM_Aufloesung_PEEM}
\end{figure*}

The spatial resolution of the photoemission electron microscope's (PEEM) imaging mode 
depends on the respective lens magnification settings. 
In the following the resolution is derived for the PEEM images shown in Figs.~\ref{fig:SOM_Aufloesung_PEEM}a and d,  reproduced from the main text. 
We extract horizontal and vertical profiles integrated over a certain width along $x$ and $y$ spatial coordinates (ranges marked by dashed orange and green lines). For the Chessy image with a field of view of \qty{800}{\micro m} diameter, a width over an equal number of sub-squares is chosen. We account thereby for
spatial variations arising from the chess-pattern-like structure that can 
lead to varying slopes of the step profile. The obtained line profile was fitted with a step-function broadened by a Gaussian to determine the width $\sigma$ of the Gaussian.
\begin{align}
f(x) &= B + \frac{A}{2}\left(\erf{\left(\frac{x-x_{0}}{\sqrt{2}\sigma}\right)} + \erf{\left(\frac{x-x_{1}}{\sqrt{2}\sigma}\right)} \right). \label{eq:error_func_fit}
\end{align}\\

The full width at half maximum ($\rm{FWHM} = 2\sqrt{2\ln(2)}\sigma$) of the Gaussian defines the spatial resolution for the respective magnification mode.

Figures~\ref{fig:SOM_Aufloesung_PEEM}b and c show the line profiles for \qty{800}{\micro\meter} field of view (Fig.~\ref{fig:SOM_Aufloesung_PEEM}a). For the large FoV lens setting we determine a spatial resolution of \qty{15.6 \pm 0.2}{} and \qty{12.5 \pm 0.1}{\micro\meter} in horizontal and vertical directions, respectively.
For the larger magnification setting with a field of view of \qty{100}{\micro m} diameter, a spatial resolution in the $x$ and $y$ directions of \qty{2.5 \pm 0.1}{\micro m} and \qty{3.0 \pm 0.1}{\micro m} was obtained. 
The latter settings demonstrate the general zooming capability of the instrument that is optimized for momentum resolution but enables the characterization of \qty{}{\micro\meter}-sized domains or sample flakes using angle-resolved photoelectron spectroscopy.

\begin{figure*}[h!]
       \includegraphics[width=17cm]{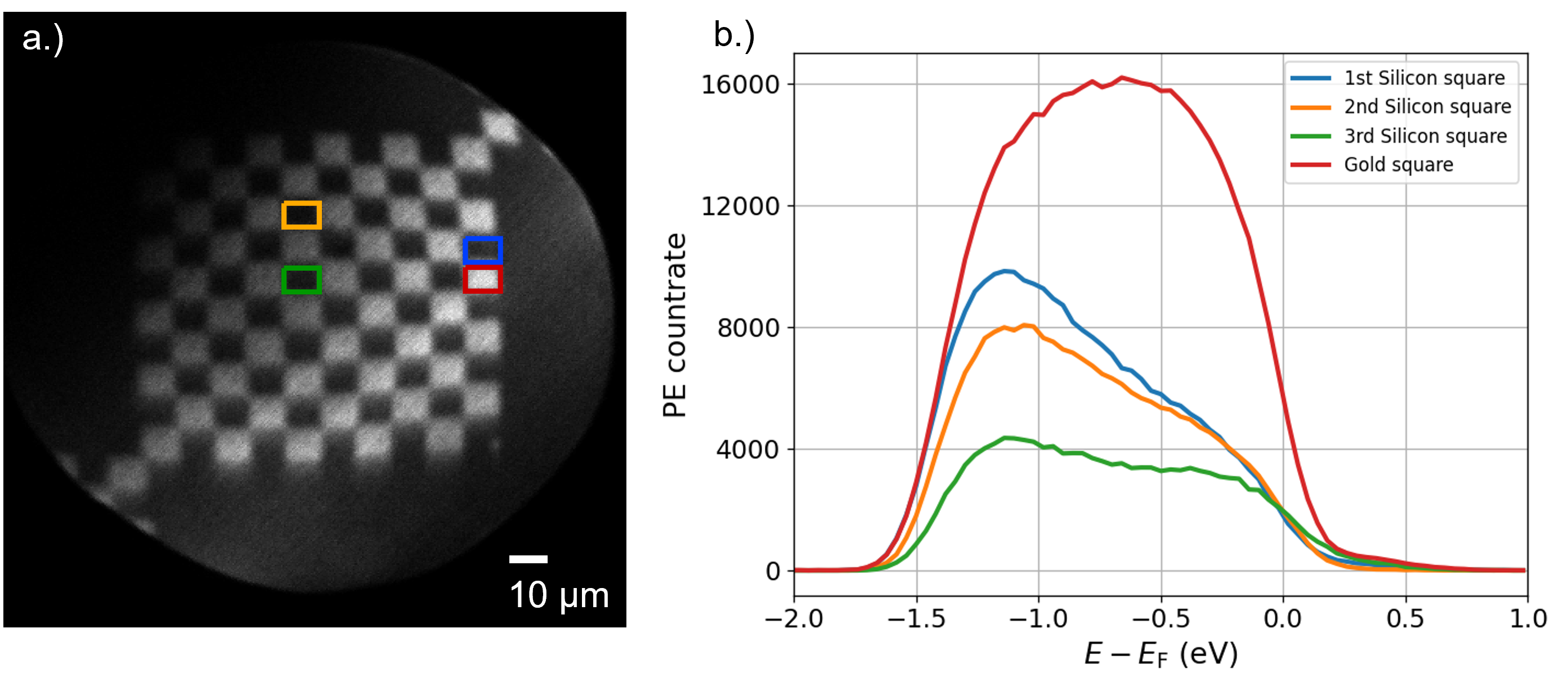}
        \caption{(a.) Spatial image of the Chessy calibration sample displaying a field of view of \qty{170}{\micro\meter} diameter. Through the illumination with a pulsed UV laser source  with \qty{220}{nm} wavelength additional energy resolution is added to the experiment. By integrating both spatial coordinates in specific areas marked by the colored rectangular boxes it is possible to monitor the photoemission intensity as a function of the energy with respect to the Fermi level ($E - E_{\rm F}$) for the $\rm SiO_{2}/Si$ and $\rm Au + SiO_{2}/Si$ squares as depicted in (b.)}
        \label{fig:Laser_chessy}
\end{figure*}

Parallel to the construction and commissioning of the PEEM, we commissioned a femtosecond Yb-based fiber laser system (central wavelength \qty{1030}{nm}) with a tuneable repetition rate up to \qty{}{MHz} (Tangerine HP, Amplitude Systems). We combined the fiber laser system with a dual-stage non-collinear optical parametric amplifier (NOPA) designed by E. Riedle (LMU).\cite{homann_octave_2008} We extended the NOPA by a second white-light stage to produce two independently tuneable fs pulses of sufficient intensity. The NOPA's broadband wavelength range of \qty{400}{nm} to \qty{930}{nm} offers variable excitation conditions for pump-probe photoemission experiments in the momentum-microscope mode. In Figs.~\ref{fig:Laser_chessy}a and b we show energy-resolved data on the Chessy calibration sample being illuminated with \qty{220}{nm} fs UV laser pulses. The UV pulses were generated through frequency doubling of the 440-nm NOPA output in a BBO crystal. They lead to direct photoemission from the calibration sample. In Fig.~\ref{fig:Laser_chessy}a, high LEY is observed for the areas patterned with gold squares ($\rm Au + SiO_{2}/Si$) whereas the low LEY squares depict bare silicon oxide ($\rm SiO_{2}/Si$) areas. We spatially select and integrate the local electron yield (marked by the colored rectangles) in order to derive the energy distribution curves shown in Fig.~\ref{fig:Laser_chessy}b. For conversion of electron time-of-flight to kinetic energy, a modified version of the Python-based work flow named "MPES" developed by the Fritz Haber institute was used.\cite{xian2020}

\newpage
\section{Difference PEEM images}
\begin{figure*}[b]
       \includegraphics[width=15cm]{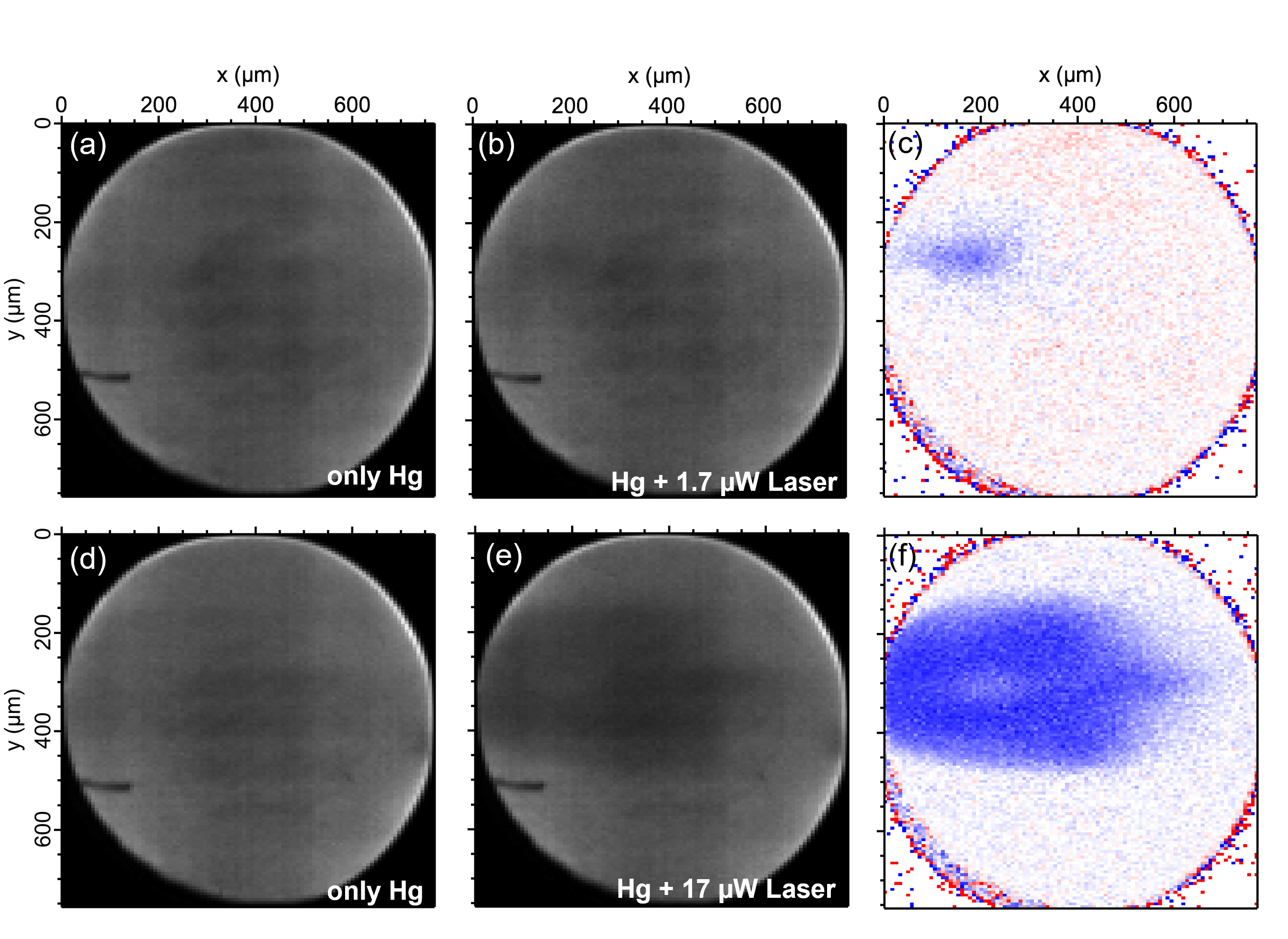}
        \caption{Spatially-resolved PEEM images showing the individual illumination conditions with only Hg lamp illumination in (a),(d) and Hg lamp + \qty{372}{nm} cw laser illumination in (b),(e). From these single raw data images the normalized difference images for the two laser fluences of \qty{1.7}{\micro\watt} and \qty{17}{\micro\watt} are derived. In the area where the combined laser+Hg lamp illumination leads to a decrease in the LEY, the highest differences with respect to single Hg lamp illumination are observed.}
        \label{fig:Diff_image_raw}
\end{figure*}
The spatially-resolved PEEM difference images shown in the main text were derived from single measurements (intensity normalized to accumulation time) with only Hg lamp illumination (cf. Fig. \ref{fig:Diff_image_raw}a, d) and combined Hg lamp + cw laser illumination (cf. Fig. \ref{fig:Diff_image_raw}b, e). From the single images, where the Hg lamp and cw laser were incident on the Az11 sample a decrease in the local electron yield is observed which translates as seen above to a local change in the work function. For the higher laser fluence a larger area on the sample is switched. The feature with a very low local electron yield on the lower left side of the PEEM image is an experimental artifact coming from a defect area of the micro-channel plate (MCP). The weak horizontal stripes originate from sensitivity variations of the MCP detector and accordingly cancel out by the normalization applied in Fig. \ref{fig:Diff_image_raw}c and f. Normalizing the difference image to the Hg PEEM image before laser illumination, leads to large differences at the boundary area of the FoV where the electron yield is very low. In these regions, we applied a threshold value to remove unwanted oversaturated pixels for the difference images shown in Fig.\,3 in the main text.